\documentclass[sigconf, nonacm]{acmart}

\setcopyright{none}
\renewcommand\footnotetextcopyrightpermission[1]{}
\usepackage{hyperref}   
\usepackage{hyperxmp}   
\usepackage{xcolor}
\usepackage{colortbl}
\usepackage{tikz}
\usepackage{balance}
\usepackage{amsmath,amsfonts}
\usepackage{graphicx}
\usepackage{capt-of}
\usepackage{booktabs}
\usepackage{varwidth}
\newsavebox \tmpbox
\usepackage[acronym]{glossaries}
\usepackage{enumitem}
\usepackage{multirow}
\usepackage{subcaption}
\usepackage{xspace}
\usepackage{makecell}
\usepackage{lscape}
\usepackage{algorithm}
\usepackage{algpseudocode}
\usepackage{amsmath}
\usepackage{multicol}
\usepackage{pifont}
\usepackage{multicol}

\newcommand{\simbox}{\textit{SIMBox}\xspace}
\newcommand{\sign}{\textit{SigN}\xspace}
\usepackage{pifont}

\usepackage{color}
\definecolor{lightred}{rgb}{1,0.6,0.6}
\definecolor{lightgray}{gray}{0.8}
\definecolor{beaublue}{rgb}{0.74, 0.83, 0.9}

\newcommand\greybox[1]{%
  \par\noindent\colorbox{lightgray}{%
    \begin{minipage}{0.47\textwidth}#1\end{minipage}%
  }%
  \vskip\baselineskip%
}

\AtBeginDocument{%
  \providecommand\BibTeX{{%
    \normalfont B\kern-0.5em{\scshape i\kern-0.25em b}\kern-0.8em\TeX}}}

\setcopyright{acmcopyright}
\copyrightyear{2018}
\acmYear{2018}
\acmDOI{XXXXXXX.XXXXXXX}

\acmJournal{POMACS}
\acmVolume{37}
\acmNumber{4}
\acmArticle{111}
\acmMonth{8}

\makeglossaries
\begin{document}

\title{\textit{SigN}: SIMBox Activity Detection Through Latency Anomalies at the Cellular Edge}

\author{Anne Josiane Kouam}
\affiliation{\institution{TU Berlin, Germany} \country{}}

\author{Aline Carneiro Viana}
\affiliation{\institution{INRIA, France} \country{}}

\author{Philippe Martins}
\affiliation{
\institution{Telecom Paris, France} \country{}}

\author{Cedric Adjih}
\affiliation{\institution{INRIA, France} \country{}}

\author{Alain Tchana}
\affiliation{
\institution{Grenoble INP, France} \country{}
}

\renewcommand{\shortauthors}{Kouam, et al.}

\begin{abstract}

Despite their widespread adoption, cellular networks face growing vulnerabilities due to their inherent complexity and the integration of advanced technologies. One of the major threats in this landscape is Voice over IP (VoIP) to GSM gateways, known as \simbox devices. These devices use multiple SIM cards to route VoIP traffic through cellular networks, enabling international bypass fraud with losses of up to \$3.11 billion annually. Beyond financial impact, \simbox activity degrades network performance, threatens national security, and facilitates eavesdropping on communications.
Existing detection methods for \simbox activity are hindered by evolving fraud techniques and implementation complexities, limiting their practical adoption in operator networks.
 
This paper addresses the limitations of current detection methods by introducing \sign, a novel approach to identifying \simbox activity at the cellular edge. The proposed method focuses on detecting \textit{remote SIM card association}, a technique used by \simbox appliances to mimic human mobility patterns. 
The method detects latency anomalies between \simbox and standard devices by analyzing cellular signaling during network attachment.
Extensive indoor and outdoor experiments demonstrate that \simbox devices generate significantly higher attachment latencies, particularly during the authentication phase, where latency is up to 23 times greater than that of standard devices. We attribute part of this overhead to immutable factors such as LTE authentication standards and Internet-based communication protocols. Therefore, our approach offers a robust, scalable, and practical solution to mitigate \simbox activity risks at the network edge.
\end{abstract}




\keywords{Cellular signaling, Network attachment, Cellular authentication}


\maketitle

\section{Introduction}
\label{sec:intro}


Cellular networks provide digital communications for more than five billion people around the globe. However, their accessibility to the general public, inherent complexity, and integration of multiple advanced technologies have exposed these networks to numerous attacks, which have significantly increased over the past decades.

In this context, Voice over IP (VoIP) to GSM gateways, also known as \simbox, are a significant source of security challenges within cellular networks. \simbox appliances bridge two telecommunication technologies by converting VoIP traffic to traditional GSM cellular networks. This allows them to route calls initiated over the internet through cellular networks by re-originating them from one of their multiple SIM cards. 

\textit{Although \simbox appliances may have legitimate uses}, such as reducing telecommunication costs or automating calls in dedicated companies, \textit{this paper highlights that their potential for misuse poses significant security risks}. Indeed, \simbox appliances are at the basis of international bypass frauds in cellular networks, recognized as one of the top four phone system frauds causing substantial losses to mobile network operators~\cite{CFCA:2021}. As depicted in Fig. \ref{fig:simbox_scheme}, International bypass fraud, or simply \simbox fraud, involves intercepting international mobile calls routing and diverting them through an internet flow (VoIP) to a \simbox in the destination country. The \simbox then re-originates the received VoIP traffic as a local mobile call from one of its SIM cards to the receiving party. 
Fraudsters bypass the regular interconnect operator, avoiding international termination charges by paying the lower local call termination charges, thus profiting from the difference.

Therefore, beyond the \textit{growing revenue loss for operators}, estimated at $2.7$ billion in 2019~\cite{CFCA:2019} and $3.11$ billion in 2021~\cite{CFCA:2021}\footnote{The CFCA’s 2023 survey summary~\cite{cfca_telecom_fraud_2023} indicates a 12\% increase in global telecom fraud losses compared to 2021, highlighting the continuing rise in the economic impact of fraud. However, the full report is not publicly available for verification.}, \simbox usage \textit{negatively impacts network quality} for legitimate consumers and \textit{compromises national security}. Specifically, \simbox fraud degrades the quality of experience for consumers due to call initiation delays and network unavailability, which in turn increases churn. Moreover, \simbox's re-originated calls introduce bias into operators' network usage records, distorting call origins and locations and affecting various analyses and research\cite{Naboulsi:2016}.
More critically, \simbox usage enables international attackers to masquerade as national subscribers, a vulnerability that could be exploited for covert operations, including by terrorists. Furthermore, \simbox appliances provide attackers with the ability to eavesdrop on international phone conversations~\cite{goantifraud:call_recording}, endangering user privacy and facilitating international espionage.



As a result, investigations into detecting \simbox activity in cellular networks have gained the attention of researchers. The objective is to provide mobile operators with means to detect and regulate \simbox usage on their networks by implementing legal registration for legitimate \simbox operations (as exemplified in \cite{t-mobile}) while blocking undeclared usage. Such investigations are typically conducted at the destination operator level, where fraud occurs. The most common approach involves analyzing network users' cellular activity traces to differentiate between \simbox patterns and legitimate ones.



Most detection methods from the literature~\cite{Sallehuddin:2013, Sallehuddin:2015, Murynets:2014, Hagos:2018, Fitsum:2020, Veloso:2020, Marah:2015} extract the spatiotemporal communication behavior of each SIM card by relying on \textit{\acrfull{CDR} traces}. 
CDRs are time-stamped and geo-referenced recordings of mobile device-generated events (i.e., call, text, data) collected by network operators. SIM cards used within \simbox appliances tend to exhibit automated behavior, distinct from human or natural patterns, characterized by low mobility, repetitive calls at odd hours, or many contacts, as noted in \cite{Murynets:2014}. 
Such literature contributions have demonstrated excellent detection performance (i.e., an average accuracy of 94.5\%) and are implemented offline, leveraging historical data collected at the network core without impacting network performance.\\
However, \simbox appliances currently available on the market offer functionalities that enable fraudsters to automatically mimic more advanced and human-like behavior, thereby evading CDR-based \simbox activity detection~\cite{Kouam:2024}. 

Conversely, a few contributions focus on the cellular edge, proposing real-time monitoring of users' network activity to detect \simbox patterns. These analyses include monitoring \textit{call audio} quality~\cite{Reaves:2015} and speakers' voices~\cite{Elrajubi:2017} to identify potential degradation due to \simbox routing. More recently, \textit{cellular signaling data} has been leveraged~\cite{Beomseok:2023} to create device model fingerprints and suggest an access-control-list prevention methodology.\\
Unfortunately, these approaches often overlook the computational challenges of cellular-edge-based deployment, which affects their practical relevance. Since these solutions must operate efficiently across the hundreds to thousands of cell towers comprising the cellular network edge, they must provide reliable indicators for detecting \simbox patterns while minimizing the computational resources needed to process them network-wide. This scalability challenge remains unresolved and explains their limited practical adoption (cf. \S \ref{sec:motivation}): e.g., \textit{call-audio}-based solutions require monitoring all local calls across the network. Similarly,
signaling-based fingerprinting necessitates maintaining an exhaustive list of device fingerprints at each base station for regular consultation.

This paper aims to bridge the gap posed by the limitations of the current solutions: \textit{It addresses the detection of \simbox patterns remaining undetected through CDR-based detection or overlooking computational costs, and proposes a novel and practical approach to unmask \simbox activities at the cellular edge.} 
The proposed approach, i.e., named \sign, identifies and leverages an indicator of \simbox activity: the \textit{remote SIM card association}.
\textit{Remote SIM card association} is a ground technology used in the \simbox to mimic human mobility pattern. It allows fraudsters to avoid the resource-intensive and easily-detected movements of \simbox appliances by enabling the binding of a SIM card to a distant gateway (with cellular antenna), as depicted in Fig. \ref{fig:simbox_architecture}. 
To the best of our knowledge, the \simbox is the only system capable of physically separating the SIM card from the cellular antenna. \textit{Remote SIM card association} is thus a distinct signature of \simbox activity resulting in decoupled network devices, unlike traditional coupled devices (e.g., phones).

\sign precisely detects such \simbox-decoupled network devices at their attachment to the network by analyzing their generated cellular signaling at the network edge. By especially characterizing the \textit{latency of devices' signaling}, we provide empirical evidence of \textit{a significant dissimilarity between \simbox-decoupled devices and coupled ones during the network attachment}. 

\noindent This includes the following contributions, outlined in Fig. \ref{fig:methodology}:
\begin{itemize}[leftmargin=*]
    \item We set up inside a Faraday shield, realistic urban settings of an operator 4G/LTE radio access and core networks with specialized equipment. Indeed, 4G is the most widespread cellular technology, particularly in the developing countries 
    where \simbox activity is the most striking, with 5G still several years away. Our testbed provides real-time access to the cellular edge network attachment signaling from 12 phones and 7 LTE \simbox appliances from two major manufacturers, collected at the base station (cf. \textbf{\S \ref{subsec:experimental_setup}}).  
    \item We make the first literature empirical characterization of network attachment latency, to the best of our knowledge (cf. \textbf{\S \ref{subsec:measurement}}). We report \simbox decoupled devices generate at least $5$ times 
    more latency than standard phones, particularly during the \textit{authentication phase} where their minimum latency is $23$ times higher. 
    \item We explain the latency overhead by analyzing the interactions between the SIM card and Mobile Equipment during the \textit{authentication phase} for a SIMBox decoupled device compared to standard devices. 
    Our investigation shows that the authentication latency in \simbox decoupled devices is influenced by unavoidable factors, such as LTE authentication standards and Internet-based communication protocols and vagaries. 
    Despite optimizations, this latency cannot match that of streamlined, legitimate devices, maintaining a clear distinction between \simbox decoupled devices and their coupled counterparts (cf. \textbf{\S \ref{subsec:explainability}}).
    \item On the other hand, we show through data collection in an actual operator network that a standard phone cannot reach such high authentication latency peaks regardless of the network signal conditions (cf. \textbf{\S \ref{subsec:tresholding}}). Our empirical findings confirm that \textit{authentication latency} is a reliable, practical, and robust metric for distinguishing \simbox decoupled devices from coupled ones.
    \item Based on this empirical evidence, we demonstrate in \textbf{\S \ref{sec:real_world_deployment}} the practicality of \sign by introducing a novel method to monitor authentication latency at the cellular edge \textit{without added overhead}. 
    Through latency distribution analysis, \sign identifies devices with \textit{consistently unusual latency values, enabling operators to promptly investigate potential threats}. Therefore, our statistical analysis shows that \sign achieves near-perfect accuracy in detecting \simbox-decoupled device attachments.
    \item To ensure the reproducibility of our results, we have released the \sign datasets and code \href{https://anonymous.4open.science/r/SigN-E485/Readme.md}{\textit{at this anonymous link}.} 
\end{itemize}
Additionally, \textbf{\S \ref{sec:background}} provides the background for our work, \textbf{\S \ref{sec:motivation}} discusses the motivation, and \textbf{\S \ref{sec:threat-model}} outlines the threat model and our defense objectives. Finally, we conclude in \textbf{\S \ref{sec:conclude}}. 
Readers can refer to the appendix for a list of acronyms used throughout the paper.

\section{Setting the stage}
\label{sec:background}
This section outlines the background for our research, covering the cellular network ecosystem and \simbox architecture.

\subsection{Cellular networks}
\vspace{0.13cm}
We overview several aspects of the 4G cellular networks, the most widespread cellular technology, particularly in the developing world where \simbox activity is the most prevalent. 

\noindent\textbf{Architecture.}  The cellular network infrastructure consists of end devices, also known as User Equipment or UE (e.g., phone, tablet), base stations, and the core network.
A User Equipment (UE) is a mobile device 
registering to the network to receive access to communication services.
It comprises two distinctive elements, the Mobile Equipment (ME) and the SIM card provided by the network operator. 
Base stations, called eNodeB in 4G networks, are intermediate connectors responsible for the radio transmission 
with the devices. At last, the core network handles administrative tasks such as the devices' authentication, security, and mobility management, intending to provide permanent service access.

\vspace{0.13cm}
\noindent\textbf{The network attachment} signaling procedure establishes a connection between end devices and the network. It occurs in four circumstances: when a device is powered on, when it moves into a new tracking area, when it loses connection with the network, or at a network trigger. In 4G, the network attachment (cf. Fig. \ref{fig:methodology}, step 2) consists of several steps aiming (i) the acquisition of the device identity, i.e., \acrfull{IMSI}, (ii) the mutual device and network authentication, (iii) the \acrfull{NAS} security setup, (iv) the device location update, and (v) the \acrfull{EPS} session establishment~\cite{3gpp_lte_attach}. 

\vspace{0.13cm}
\noindent\textbf{The SIM card} binds the mobile subscription to the network device.
It securely stores the \acrshort{IMSI} subscriber identifier and a secret symmetric key called the subscriber key (or $K_i$, in short) used in steps (ii) and (iii) of the network attachment. It also represents an environment protected from attackers where the network authentication and security algorithms are run following the \acrshort{AKA} protocol~\cite{3gpp_aka}.

\subsection{\simbox architecture and fraud}

\begin{figure}
\centering
\includegraphics[width=0.9\linewidth]{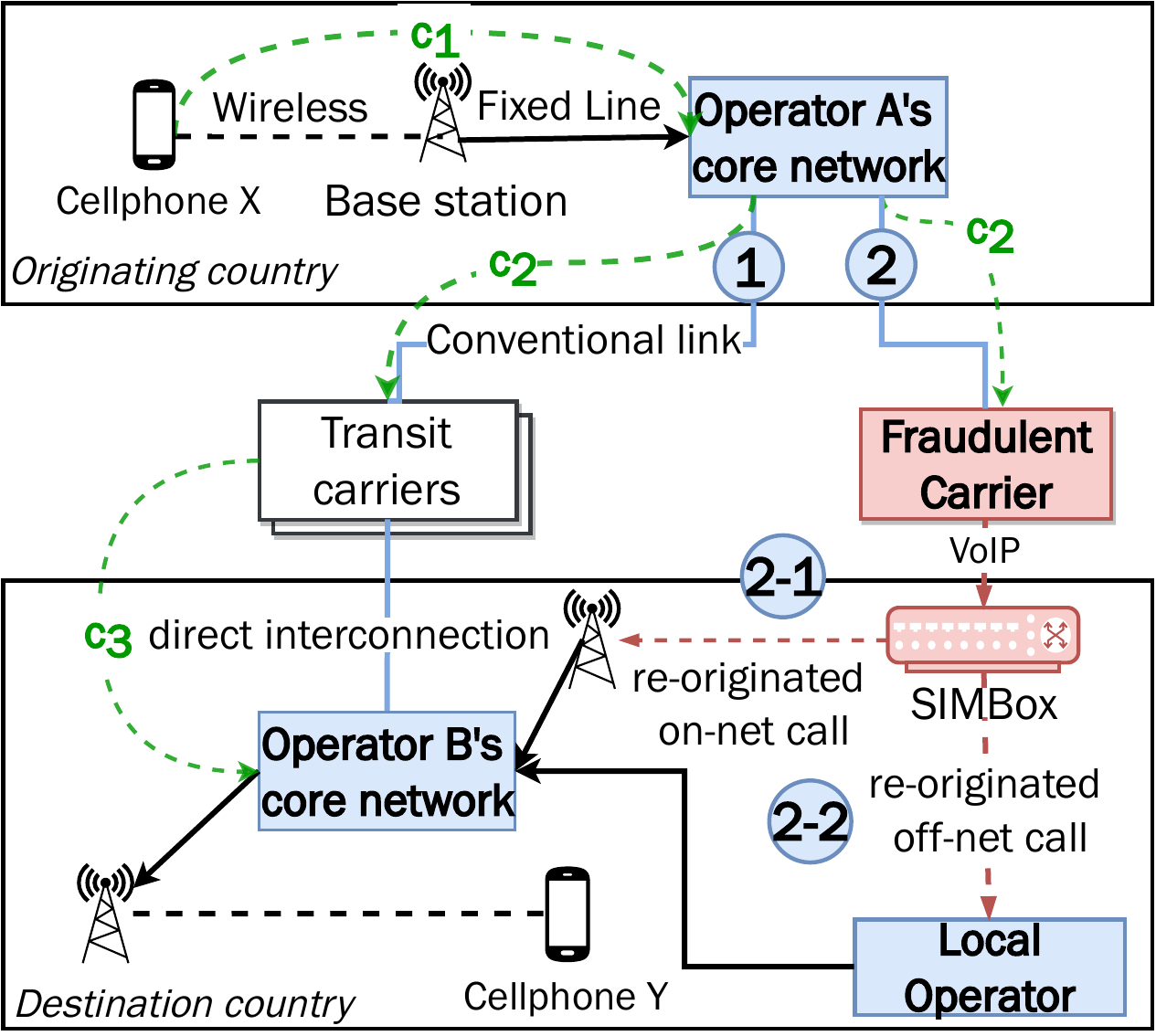}
\caption{International call routing: (Flow 1) Legitimate scheme, (Flow 2) Fraudulent scheme.}
\label{fig:simbox_scheme}
\end{figure}

\vspace{0.13cm}
\noindent\textbf{\simbox fraud scheme.} As depicted in Fig. \ref{fig:simbox_scheme}, \simbox fraud interferes with the regular international voice call routing (\textit{flow 1}). 
In a regular routing, the call traffic leaves the caller's mobile operator (Operator A) and is routed to the destination country through a set of transit operators. The traffic is received directly by the called party's operator (Operator B), who terminates it.
Nevertheless, a transit carrier can be fraudulent. Indeed, transit carriers perform traffic interconnection between countries by buying and reselling international termination routes. A fraudulent carrier instead diverts the traffic it receives through a low cost VoIP trunk, as in the \textit{flow 2} on Fig. \ref{fig:simbox_scheme}.
The diverted traffic is sent to a \simbox (VoIP to GSM gateway) located in the destination country and re-originated as a national mobile call to its recipient. 
Once in the destination country, there are two possible fraudulent termination scenarios: (i) 2-1 
is an on-net termination when the re-originated call is made using a SIM card of Operator B, the same operator of the called party, and (ii) 2-2 is an off-net termination when the fraudster uses a SIM card from a different local operator in the destination country.

\vspace{0.13cm}
\noindent\textbf{The \simbox} operates 
as a VoIP GSM gateway. It receives a diverted call traffic as a VoIP client and terminates it by re-originating a  cellular mobile call using one of its numerous SIM cards. 
The \simbox continuously creates network devices by associating SIM cards and GSM modules (providing wireless link to the network). 
The \simbox includes three kinds of hardware components:
\begin{itemize}[leftmargin=*]
    \item The \textit{gateway} is a rack with a set of GSM modules maintaining the wireless communication inside a given cellular frequency range 
    (i.e.,2G/3G/4G). It receives incoming VoIP traffic and distributes it to the GSM modules 
    for termination as mobile calls. The gateway plays the role of Mobile Equipment for the formed \simbox devices. 
    Hence, the recorded network location of a \simbox device is the location of its belonging gateway.
    Most gateways in the market include SIM slots for operation. 
    \item The \textit{SIMBank} is an appliance with numerous SIM slots that remotely holds a bundle of SIM cards (e.g., 128 in the SMB128 model~\cite{SMB128}). It manages \simbox SIM cards, including their addition, removal, and data transfer. 
    \item The \textit{control server} is a web server providing the \simbox control functions, i.e., binding of SIM cards to GSM modules and architecture configuration. It can be hosted online to ease remote access from a web client.
\end{itemize}

\begin{figure}
\centering
\includegraphics[width=\linewidth]{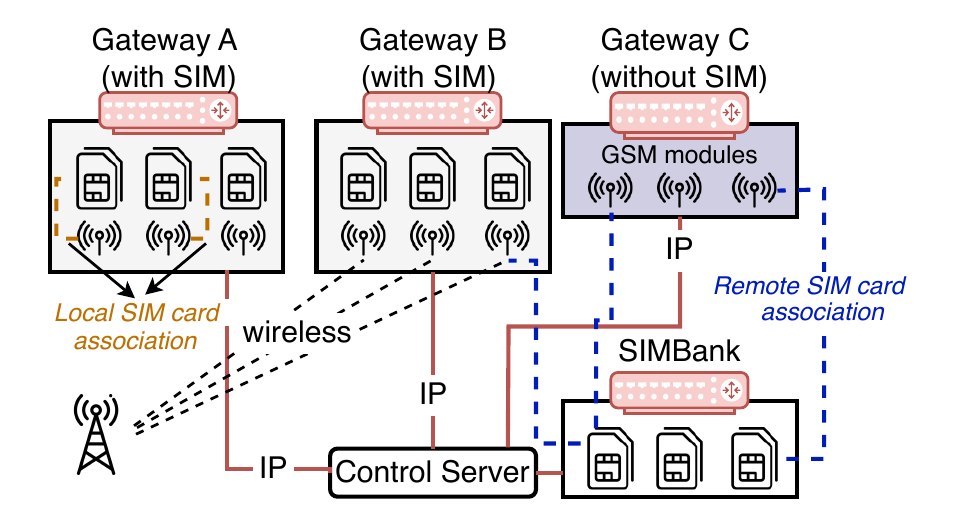}
\caption{Example of a \simbox distributed architecture.}
\label{fig:simbox_architecture}
\end{figure}

Distributed \simbox architecture involves the interaction of such appliances over an IP network using TCP or UDP protocols. Hence, 
as shown in Fig. \ref{fig:simbox_architecture}, \simbox devices formation can be done through \textit{local SIM card association} if the SIM card is in the same appliance as its associated GSM module, or \textit{remote SIM card association} if the SIM card is from another appliance, i.e., the SIMBank. \textit{Local SIM card association} results in \textit{coupled UEs}, while \textit{remote SIM card association} yields \textit{decoupled UEs}.


\section{Unraveling literature gaps}
\label{sec:motivation}
Despite the significant impact of \simbox activity, it has received limited attention in the literature, with only 15 detection methods proposed since 2011. We categorize these contributions based on the type of cellular network data they use and how it is processed.
Specifically, we distinguish between methodologies that rely on core network data (e.g., Call Detail Records/CDRs, \S \ref{subsec:offline}) and those that use edge network data (e.g., call audio and cellular signaling, \S \ref{subsec:online}). 
This section highlights the strengths and weaknesses of current \simbox detection approaches and positions \sign to address the remaining gaps. 
For a complete survey of \simbox fraud solutions before 2021, refer to \cite{Kouam:2021}.

\subsection{Network-core-based detection}
\label{subsec:offline}


Literature approaches operating at the network core~\cite{Sallehuddin:2013, Sallehuddin:2015, Murynets:2014, Hagos:2018, Fitsum:2020, Veloso:2020, Marah:2015} aim to identify \simbox activity 
from SIM cards' communication and mobility behavior extracted from CDRs datasets. These methods rely on various features (e.g., \#calls at night, \#contacts, \#incoming calls, \#locations) to distinguish SIM cards used for \simbox termination from SIM cards used by genuine consumers. \textit{Such contributions have demonstrated high accuracy, averaging 94.5\%, in detecting \simbox activity characterized by unusual patterns in communication or mobility, as in \cite{Marah:2015, Murynets:2014, Kehelwala:2015}: numerous outgoing calls, few stay points, SIM card clusters, or no incoming calls.}

Unfortunately, \simbox devices have evolved with functionalities that automatically mimic human behavior in CDR datasets, known as \acrfull{HBS}~\cite{Kouam:2021}. \acrshort{HBS} techniques enable \simbox devices to operate while maintaining human-like behavior in terms of communication and mobility. In communication, this is achieved by thresholding the number and duration of initiated calls and controlling their contacts and timing. For mobility, fraudsters use \textit{remote SIM card association}, binding a SIM card to a remote gateway (i.e., Mobile Equipment), resulting in an erroneous network recording of the SIM card location. 
Notably, the automatic binding of a \simbox SIM card to gateways in different locations at various times creates human-like movements between network cells in CDRs at no cost to fraudsters.

Recent research~\cite{Kouam:2024} empirically examines the performance of \acrshort{HBS}-generated \simbox patterns compared to CDR-based \simbox activity detection. The results indicate that the current \acrshort{HBS} functionalities produce \simbox patterns that closely mimic human behavior, enabling them to evade detection by CDR-based methods with a high degree of success. 
\textit{This finding underscores the limitations of CDR-based approaches, which are insufficient to unmask all existing \simbox patterns.}

\vspace{0.13cm}
\greybox{
\noindent \textbf{Insight:} \textit{The wide adoption of \acrshort{HBS} techniques in the \simbox ecosystem limits the effectiveness of existing network-core-based \simbox activity detection, justifying the need for detection techniques tailored to the fraud evolution.}}

\subsection{Network-edge-based detection}
\label{subsec:online}

Network-edge-based detection methods operate at each base station within the cellular network, monitoring activity in real-time to detect \simbox patterns. Unlike network-core-based solutions, these methods often face scalability challenges that affect their practical efficiency. Specifically, existing solutions analyze either \textit{call audio} or \textit{cellular signaling data}.

First, \textit{call audio-based} solutions, which examine speakers' voices~\cite{Elrajubi:2017} or call quality~\cite{Reaves:2015}, have demonstrated efficiency in lab settings but face real-world deployment challenges. Investigating all local calls across the network raises privacy concerns and significant scalability issues.

\begin{figure}
\begin{subfigure}{0.49\textwidth}
\centering
\includegraphics[scale=0.39]{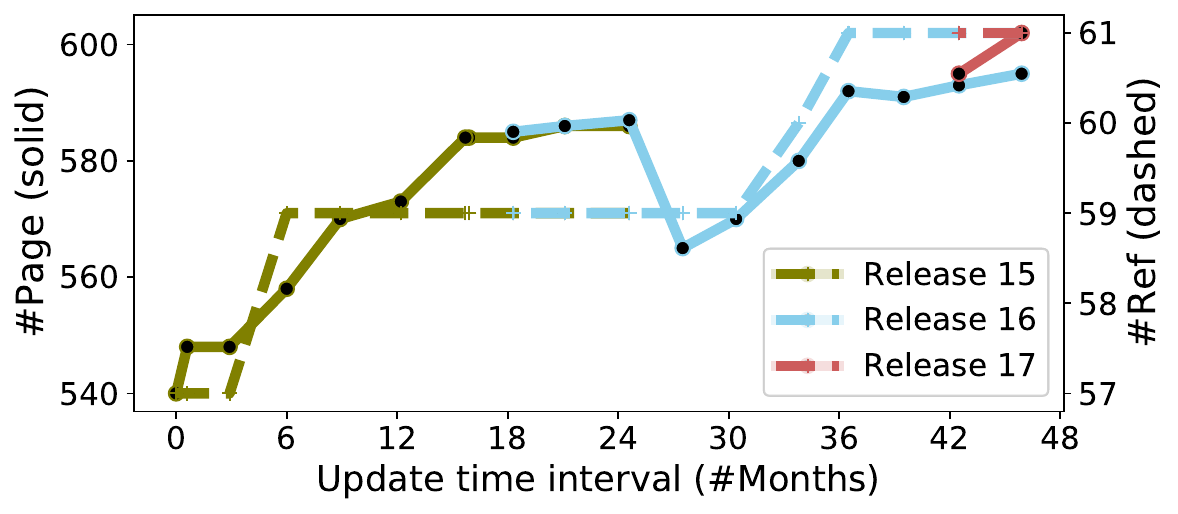}
\caption{NAS spef. updates from 27-03-2019 to 03-01-2023.} 
\label{fig:nas_updates}
\end{subfigure}
\hfill
\begin{subfigure}{0.5\textwidth}
\centering
\includegraphics[scale=0.39]{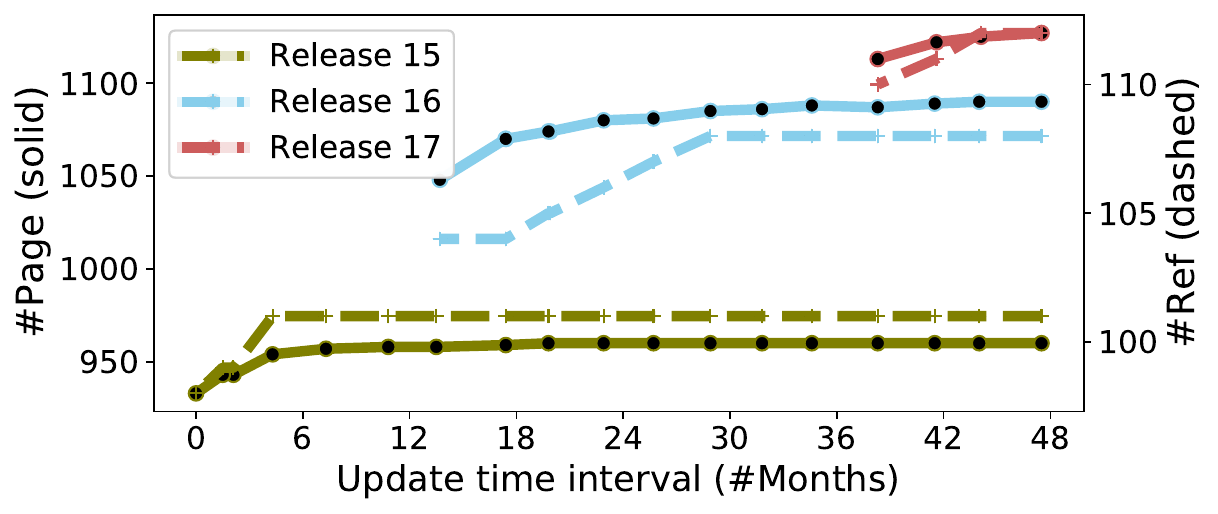}
\caption{RRC spef. updates from 19-02-2019 to 13-01-2023.} 
\label{fig:rrc_updates}
\end{subfigure}
\caption{NAS\cite{NAS_3gpp} and RRC\cite{RRC_3gppp} protocol specifications size. Number of pages (\#Page) in solid line. Number of references (\#Ref) in dashed line.}
\label{fig:spef_updates}
\vspace{-0.5cm}
\end{figure}

In contrast, Oh et al. \cite{Beomseok:2023} leverages \textit{cellular signaling data} for \simbox detection, proposing a fingerprinting-based approach, referred to as \textit{ACLPrint}. This method compares the device's fingerprint and factory identifier code (i.e., \acrshort{TAC}) to a pre-established database, rejecting devices with mismatched fingerprints.
However, \textit{ACLPrint} faces significant scalability issues. Frequent updates to the relying 3GPP LTE specifications (cf. Fig. \ref{fig:spef_updates}), which occur roughly every three months, require constant manual monitoring of extensive documents and their numerous references, along with adjustment of the fingerprinting process. Additionally, \textit{ACLPrint} relies on a pre-established database, making it vulnerable to new, unrecorded \acrshort{ME} models and brute-force attacks, where fraudsters modify their \simbox identifiers until they find a bypass. This also implies each network base station maintains and regularly consults such a vast database, complicating deployment.

These findings highlight the scalability challenges of \textit{ACLPrint} and similar methods, limiting their real-world efficiency.
\vspace{0.13cm}
\greybox{
\noindent \textbf{Insight:}  \textit{The effectiveness of network-edge-based detection methods hinges on their scalability — their capacity to function across the entire network without compromising network performance. However, this challenge remains unmet in current literature contributions.}}

\section{Establishing \sign grounds}
\label{sec:threat-model}


In this section, we overview our investigation standpoint. In \S \ref{subsec:threat_model}, we first outline the \sign threat model, defense objectives, and key insights. Then, in \S \ref{subsec:preliminaries}, we explain why signaling latency is central to the \sign methodology. Further discussions in \S \ref{subsec:network_attachment_focus} provide an overview of signaling procedures in LTE, justifying \sign focus on the network attachment procedure signaling.



\subsection{Threat models and defense goals}
\label{subsec:threat_model}

\vspace{0.13cm}
\noindent\textbf{Threat model:}
\sign is designed in a complementary viewpoint to literature methodologies. It addresses "advanced \simbox activity" that is undetectable with CDR-based approaches due to the use of \acrfull{HBS} (cf. \S \ref{subsec:offline}), and with network-edge-based methods due to privacy/scalability limitations (cf. \S \ref{subsec:online}).

Therefore, we focus on detecting advanced \simbox patterns derived from HBS techniques implementation. As described in \S \ref{subsec:offline}, these patterns involve \textit{remote SIM card associations} to mimic human-like mobility behavior. We, therefore, assume that adversaries implement \textit{remote SIM card association} using a distributed \simbox architecture (cf. Fig \ref{fig:simbox_architecture}) with lawfully-issued local SIM cards held within a SIMBank.


\vspace{0.13cm}
\noindent\textbf{Defense objective:} 
Our goal is the efficient, online prevention of such advanced \simbox activity in a privacy-preserving and practical manner. We introduce \sign, \textit{a network-edge-based \simbox activity detection methodology based on cellular signaling data}. \sign aims to prevent fraudulent \simbox activity on the mobile network surface. This is done through the reliable identification of network devices with advanced \simbox patterns at the cellular edge \textit{before any fraudulent calls} are made. Such an early detection effectively \textit{prevents \simbox owners from gaining financial advantage}. \textit{\sign thus provides mobile operators with the means to detect and regulate \simbox usage, enforcing legal registration for legitimate operations while blocking undeclared usage.
}

We design \sign keeping in mind the open challenges of the network-edge-based \simbox fraud detection literature (cf. \S \ref{subsec:online}). To ensure high real-world relevancy, we establish the following requirements: (i) \textit{Privacy}: \sign should rely on network device features that operators can access without impeding privacy. (ii) \textit{Practicality}: \sign implementation should be scalable and require minimal, non-constant effort from operators for wide deployment on the network surface.


\vspace{0.13cm}
\noindent\textbf{Key insights:} 
As advanced \simbox patterns result from carefully crafted communications and movements to mimic human behavior, providing precise online detection indicators for mitigation at the cellular edge is a genuine challenge.

Our approach to addressing this challenge builds upon the indicator of advanced \simbox activity: \textit{remote SIM card association}. Specifically:
\begin{enumerate}[leftmargin=*]
\item \textit{Indispensability of \textit{remote SIM card association}}: \textit{Remote SIM card association} is essential for \simbox to mimic human behavior (cf. \S \ref{subsec:offline}). Previous work~\cite{Kouam:2024} establishes that \simbox activity without \textit{remote SIM card association} results in distinctive communication behaviors efficiently detected through existing methods. 
\item \textit{Uniqueness of \textit{remote SIM card association} to \simbox}: No other user end device (smartphones, tablets, laptops, IoT devices, modems, etc.) separates the SIM card from the \acrfull{ME} during network operations, making \textit{remote SIM card association} an explicit proxy for advanced \simbox activity.
\end{enumerate}

Henceforth, by detecting the use of \textit{remote SIM card association}, \sign effectively controls advanced \simbox activity and prevents any malicious usage.
\sign analyzes cellular signaling to determine if an attaching device is either \textit{coupled} or \textit{decoupled} via a \simbox-operated \textit{remote SIM card association}.
This approach follows the intuition that signaling messages from \simbox-decoupled devices exhibit higher latency compared to coupled devices, as below.


\subsection{Preliminaries}
\label{subsec:preliminaries}

\begin{figure}
    \centering
    \includegraphics[width=0.6\linewidth]{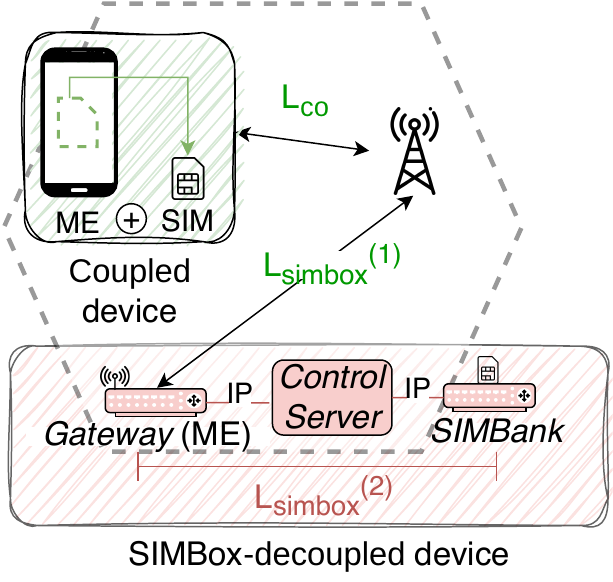}
    \caption{Signaling latency of coupled and \simbox-decoupled devices.}
    \label{fig:sig_latency}
    \vspace{-0.7cm}
\end{figure}

Standard devices in cellular networks are a combination of a \acrfull{ME} and a SIM card integrated within the \acrshort{ME}, as depicted in Fig. \ref{fig:sig_latency}: \textit{it is a coupled ME-to-SIM combination}. 
They thus present a coupled signaling latency $L_{co}$, corresponding to the interaction time between the base station and the \acrshort{ME}. \

On the other hand, \simbox-decoupled devices make a \textit{logical IP-based} binding of a GSM module (inside the gateway operating as the \acrshort{ME}) to a SIM card (inside the SIMBank) done at the level of the \textit{control server}: \textit{it is a decoupled ME-to-SIM combination.} 
Accordingly, the signaling latency of a \simbox-decoupled device, i.e., $L_{simbox}$, includes:
\begin{itemize}
    \item $L_{simbox}^{(1)}$ corresponding to the interaction time between the base station and the gateway (i.e., \acrshort{ME} of the \simbox-decoupled device), and
    \item $L_{simbox}^{(2)}$ corresponding to the interaction time between the gateway (i.e., ME) and the \simbox-decoupled device's SIM card inside the SIMBank
\end{itemize}
\noindent such that $L_{simbox} = L_{simbox}^{(1)} + L_{simbox}^{(2)}$.

\vspace{0.2cm}
Therefore, compared to coupled devices' signaling latencies $L_{leg}$, \simbox-decoupled devices' signaling latency $L_{simbox}$,  will tend to be larger due to component $L_{simbox}^{(2)}$ involving one or more exchanges of \simbox components over the Internet during the signaling operation.



\sign methodology aims to identify such latency overhead, at the base station, to distinguish between coupled devices and \simbox-decoupled ones. It, therefore, fulfills our mitigation requirements as follows: 
\begin{enumerate}[leftmargin=*]
\item Privacy: Mobile operators have a natural access to signaling latency measurements as these do not relate to any specific individual or device model's identifier or data content and, thus, are not privacy-impeding. 
\item Practicality: Inspecting signaling latency is straightforward and is already implemented in LTE using \textit{timers}. This practical approach ensures ease of deployment at no additional cost across the entire network edge.
\end{enumerate}

\begin{table}
\centering
\begin{minipage}{\linewidth}
\caption{Summary of signalling procedure analysis}
\label{tab:signaling-proc}
\resizebox{\columnwidth}{!}{%
\begin{tabular}{l|l|l|l}
\hline
\textbf{\begin{tabular}[c]{@{}l@{}}Signaling \\ procedure\end{tabular}}    & \textbf{Description}                                                                                                 & \textbf{\begin{tabular}[c]{@{}l@{}}\#device\\ processing\end{tabular}} & \textbf{\begin{tabular}[c]{@{}l@{}}Moment of occurrence\end{tabular}}                                                                                                        \\ \hline
\begin{tabular}[c]{@{}l@{}}Network \\ attachment\end{tabular}              & \begin{tabular}[c]{@{}l@{}}Device connection and \\ authentication to the \\ network\end{tabular}                    & 4                                                                      & \begin{tabular}[c]{@{}l@{}}- At the device power on\\ - Device mobility dependent\\ - Network initiated\end{tabular} \\ \hline
Handover (X2)                                                              & \begin{tabular}[c]{@{}l@{}}Direct device connection's\\ transfer between network\\ base stations\end{tabular}        & 1                                                                      & Device mobility dependent                                                                                            \\ \hline
Handover (S1)                                                              & \begin{tabular}[c]{@{}l@{}}Device connection's transfer\\ between base stations via \\ the core network\end{tabular} & 1                                                                      & Device mobility dependent                                                                                            \\ \hline
CQI update                                                                 & \begin{tabular}[c]{@{}l@{}}Device's information to the\\  network of channel quality\end{tabular}                    & 0                                                                      & \begin{tabular}[c]{@{}l@{}}Network dependent \\ (periodic or aperiodic)\end{tabular}                                 \\ \hline
\begin{tabular}[c]{@{}l@{}}Data bearer \\ establishment\end{tabular}       & \begin{tabular}[c]{@{}l@{}}Setting up data\\  transmission channel\end{tabular}                                      & 2                                                                      & \begin{tabular}[c]{@{}l@{}}Device communication \\ dependent\end{tabular}                                            \\ \hline
\begin{tabular}[c]{@{}l@{}}Mobile originated \\ SMS signaling\end{tabular} & \begin{tabular}[c]{@{}l@{}}Texts transmission from\\ the device to the network\end{tabular}                          & 1                                                                      & \begin{tabular}[c]{@{}l@{}}Device communication\\ dependent\end{tabular}                                             \\ \hline
\begin{tabular}[c]{@{}l@{}}CS Fallback\\ call setup\end{tabular}      & \begin{tabular}[c]{@{}l@{}}Establishment of a\\ traditional voice call circuit\end{tabular}                          & 7                                                                      & \begin{tabular}[c]{@{}l@{}}Device communication \\ dependent\end{tabular}                                            \\ \hline
\end{tabular}%
}
\end{minipage}
\end{table}

\subsection{Focusing on the network attachment}
\label{subsec:network_attachment_focus}

LTE standards provide several signaling procedures to deliver communication services to network devices.
Aiming to distinguish \simbox-decoupled devices by their latency overhead, we analyze the most common of such signaling procedures (cf. Table \ref{tab:signaling-proc}) based on two criteria:
First, their \textit{ability to involve device's processing}, i.e., the number of device processing necessarily occurring during the signaling procedure (\textit{\#device processing}), that maximizes the latency checking possibilities to detect latency anomalies of \simbox-decoupled devices; 
Second, their \textit{moment of occurrence} indicating when and how often a latency checking can done and whether such checking depends on the network or on the device behavior. 

Our investigation relies upon the related 3GPP specifications and 
reports in Table \ref{tab:signaling-proc} the uncovered \textit{\#device processing} and \textit{moment of occurrence} per signaling procedure. We make the following observations:
\vspace{-0.1cm}
\begin{itemize}[leftmargin=*]
    \item The number of device processing varies from one procedure to another, indicating that procedures with greater values are more suitable for our \simbox activity detection goal. For instance, CQI updates, though fully network-controlled, do not involve any device processing, therefore, not allowing to uncover \simbox latency overhead.
    \item Concerning the moment of occurrence, signaling procedures are triggered either by the device's communication or mobility behavior or by the network itself. Network-initiated or mandatory procedures are more relevant to guarantee minimal interference by fraudsters. For instance, 
    data bearer establishment are only executed when a device starts a mobile data session, that can be avoided by fraudsters. 
\end{itemize}

Based on these insights, \textit{the network attachment procedure is the optimal choice for building \sign}, as it incurs a sufficient number of device processing compared to other signaling procedures. This procedure is mandatorily carried out by all network devices (coupled or \simbox-decoupled) when they connect to an operator network upon being powered on.
It, therefore, enables the implementation of a  network access control that prevents any \simbox activity-induced damage. Furthermore, it can be triggered by the operator (as a Tracking Area Update), independently of the device's behavior, to increase attempts to detect \simbox activity.

In the following steps, we carry out an in-depth empirical study of the network attachment signaling latency to assess if this metric is satisfactory in distinguishing between coupled and \simbox-decoupled devices. 

\section{Latency empirical study}
\label{sec:latency_detection}


\begin{figure*}
  \centering
  \includegraphics[width=\linewidth]{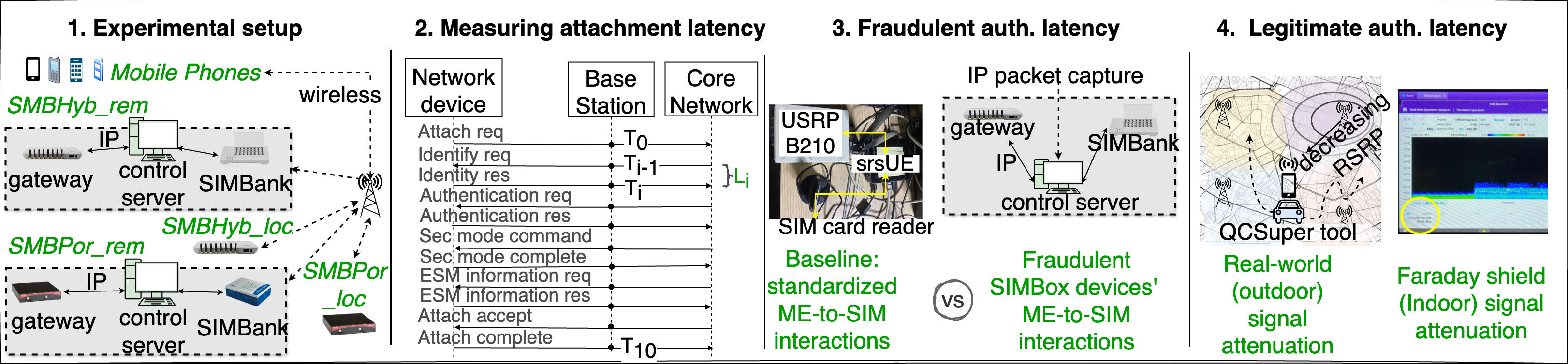}
  \caption{\sign attachment latency analysis methodology.}
  \label{fig:methodology}
\end{figure*}

In this section, we want to evaluate how well the network attachment signaling latency (referred to as \textit{attachment latency}, for simplicity) can be used to differentiate  between coupled and \simbox-decoupled devices through experimental studies. To this end, we answer the following questions: 
\begin{enumerate}
  \item[$\textbf{[Q1]}$] \textit{How different is the attachment latency between coupled and \simbox-decoupled devices?} 
  \item[$\textbf{[Q2]}$] \textit{What factors explain the attachment latency of \simbox-decoupled devices, and is this latency reliable?}
  \item[$\textbf{[Q3]}$] \textit{What factors influence the attachment latency of coupled devices, and how might these variations compare to the latency observed in \simbox-decoupled devices?}
\end{enumerate}

Through extensive indoor and outdoor experiments, represented in Fig. \ref{fig:methodology}, we make important observations relatively to the previous questions, which are summarized as follows: 

\begin{enumerate}
  \item[$\textbf{[O1]}$] \textit{\simbox-decoupled devices generate at least $5\times$ more attachment latency compared to coupled ones 
  in particular during the \textit{authentication phase} where their minimum latency is  $23\times$ higher (cf. \S \ref{subsec:measurement}). }
  \item[$\textbf{[O2]}$] \textit{\simbox-decoupled devices' latency is induced by both (i) \simbox implementation and (ii) network protocols-imposed procedures (i.e., LTE standards and TCP/UDP correction and retransmission mechanisms). While the former allows for fraudsters improvement, the latter is beyond fraudsters reach and guarantee a minimum latency still $2\times$ higher than coupled devices, in the authentication phase} (cf. \S \ref{subsec:explainability}).
  \item[$\textbf{[O3]}$] \textit{Regardless of the wireless network channel conditions, coupled devices' latency, in the authentication phase, cannot reach high values comparable to the one of \simbox-decoupled devices (cf. \S \ref{subsec:tresholding}). Therefore, the authentication latency \textit{unambiguously separates} coupled and \simbox-decoupled devices.}
\end{enumerate}


\subsection{Experimental Setup}
\label{subsec:experimental_setup}

To ensure our study complies with regulations and avoids interference with live operator networks, we have designed a high-performance testbed that accurately simulates a real-world 4G cellular network. This setup utilizes Amarisoft's professional suite, a trusted solution in the wireless industry, and is housed within a  $30m^2$ Faraday shield. Amarisoft's software-based technology is fully compliant with 3GPP standards and compatible with off-the-shelf hardware, including the physical layer~\cite{amarisoftWebsite}. With over 1,000 customers in more than 60 countries, including numerous public and private network operators, Amarisoft's solutions are widely adopted for both laboratory and field applications. This widespread adoption underscores the reliability and accuracy of Amarisoft's technology in replicating authentic network environments. Consequently, the signaling we capture in our testbed mirrors the exact procedures employed in operational networks, ensuring that our results are both precise and reflective of real-world conditions.

Our testbed employs a single PC to run both the base station and core network nodes, including the \acrshort{MME}, \acrshort{IMS}, and \acrshort{SGW}. This PC handles baseband processing, while radio processing is managed by a software-defined \acrshort{USRP} B210 connected to the PC, enabling seamless integration of the baseband and SDR systems. Detailed specifications of all testbed components, including the featured 4G cell and its radio parameters, are provided in Table \ref{tab:testbed} in the appendix. Signal quality, specifically \acrshort{RSRP}, has been rigorously validated using a radio spectrum analyzer, showing excellent performance (around -71 dBm) consistent with real-world urban network conditions as documented in recent studies~\cite{Krawczeniuk:2019} (cf. Fig. \ref{fig:spectrum0}).

Our setup includes 12 mobile phones from five different vendors and 7 \simbox devices from two manufacturers—Hybertone, the leader in the \simbox market~\cite{goantifraud_top5}, and Portech. These devices are equipped with programmable SIM cards~\cite{sysmocom}, ensuring they connect seamlessly to the LTE network inside the shield. Notably, Hybertone \simbox appliances support both TCP and UDP protocols, while Portech devices only support UDP. 
As illustrated in Fig. \ref{fig:methodology}, step 1, we use the \simbox appliances of both manufacturers to deploy (i) \textit{remote SIM card association} and (ii) \textit{local SIM card association}. 
The \textit{remote SIM card associations} a \simbox control server hosted on LAN-connected PCs (cf. Table \ref{tab:testbed}).
Thus, \textit{SMBHyb\_rem} and \textit{SMBPor\_rem} refer to devices using \textit{remote SIM card association} from Hybertone and Portech, respectively, which are \simbox-decoupled. Likewise, \textit{SMBHyb\_loc} and \textit{SMBPor\_loc} refer to devices using \textit{local SIM card association} from the same manufacturers. These coupled devices, though potentially fraudulent, fall outside the scope of this research, as they involve already-addressed \simbox activity (cf. \S\ref{subsec:threat_model})

\subsection{Measuring attachment latency}
\label{subsec:measurement}

Here we collect and analyze, for all the coupled and \simbox-decoupled devices of our testbed, the latency at each step of network attachment procedure (cf. Fig. \ref{fig:methodology}, step 2). 
We first detail the methodology for latency collection and computation and then present and discuss the obtained results.

\vspace{0.13cm}
\noindent\textbf{Methodology.}
For each network device (i.e., phone model, \simbox coupled and decoupled devices), we carry out 50 executions of the network attachment procedure. The resulting cellular signaling logs are recorded at the level of the base station.
We consider only NAS-layer logs, as they provide information on the signaling between the device and the core network during the network attachment. These logs consist of 11 messages, as represented in Fig. \ref{fig:methodology}, step 2. 
For each message we use the following associated fields for latency computation:  
\verb|time, layer, direction, device_id, message|. 
The communication direction, i.e., uplink or downlink, indicates the message originator as the device or the network, respectively. 
Therefore, we compute the latency of each message as $L_i = T_i - T_{i-1}$, i.e, the delta time between the arrival of a message $i$ and its previous one $i-1$. Depending on the message direction (uplink/downlink), $L_i$ refers to the network's or the device’s processing time along with the message transmission time to the base station. The total network attachment latency of a network device is thus $\sum_{i=1}^{n} T_i - T_{i-1}$ with $n=10$ steps (cf. Fig. \ref{fig:methodology}, step 2).

\vspace{0.13cm}
\noindent\textbf{Results.}
Table \ref{tab:detailed_latency} reports the obtained latency's mean and standard deviation values for each step of the network attachment. A particular interest is on the lines with \textit{uplink} direction, 
enabling us to determine and compare the processing time per network device. We make the following observations:
\begin{itemize}[leftmargin=*]
  \item  Regardless of the model, all phones have comparable latencies per step, similar to the \simbox devices resulting from \textit{local SIM association}. However, \textit{distinguishing from coupled devices, the attachment latency for \simbox-decoupled devices 
  is significantly higher i.e., $\approx 9\times$ for Hybertone and $\approx 5\times$ times for Portech.}
  \item \textit{Such latency distinction of \simbox-decoupled devices emerges at step 4 (authentication response)}, which consists of mutual authentication of the network and device, following the \acrshort{AKA} procedure~\cite{3gpp_aka}. 
  The authentication phase involves a computation internal to the SIM card (in the remote SIMBank) and, therefore, necessarily imputes IP-based interactions between \simbox components, explaining the overhead. Particularly in the authentication phase, \simbox-decoupled devices show a latency approximately $29\times$ (for Hybertone) and $23\times$ (for Portech) higher than coupled devices' latency.
\end{itemize}

\greybox{\noindent\textbf{Insight.} \textit{The previous results spotlight the authentication phase as the primary source of latency distinction for \simbox-decoupled devices during the network attachment. Henceforth, we narrow the following investigations to 
understand such authentication latency.}}
\vspace{-0.5em}

\begin{table*}[]
\centering
\caption{Latency (in ms) per device model reported per network attachment step}
\label{tab:detailed_latency}
\resizebox{\textwidth}{!}{%
\begin{tabular}{|ll|l|l|l|l|l|l|l|l|l|l|>{\columncolor{lightred}}l|l|>{\columncolor{lightred}}l|}
\hline
\multicolumn{1}{|l|}{\textbf{Step}} &
\textbf{Direction} &
\textbf{\begin{tabular}[c]{@{}l@{}}\rotatebox[origin=c]{90}{\makecell{Fair\\Phone5G}}\end{tabular}} &
\textbf{\rotatebox[origin=c]{90}{\makecell{Galaxy\\A90}}} &
\textbf{\rotatebox[origin=c]{90}{\makecell{Galaxy\\Note4}}} &
\textbf{\rotatebox[origin=c]{90}{\makecell{Galaxy\\S3}}} &
\textbf{\begin{tabular}[c]{@{}l@{}}\rotatebox[origin=c]{90}{\makecell{Galaxy\\ZFold25G}}\end{tabular}} &
\textbf{\begin{tabular}[c]{@{}l@{}}\rotatebox[origin=c]{90}{\makecell{OnePlus\\Nord}}\end{tabular}} &
\textbf{\begin{tabular}[c]{@{}l@{}}\rotatebox[origin=c]{90}{\makecell{Sony\\XPERIA}}\end{tabular}} &
\textbf{\begin{tabular}[c]{@{}l@{}}\rotatebox[origin=c]{90}{\makecell{Xiaomi10\\Lite5G}}\end{tabular}} &
\textbf{\begin{tabular}[c]{@{}l@{}}\rotatebox[origin=c]{90}{\makecell{Xiaomi9\\Pro5G}}\end{tabular}} &
\textbf{\begin{tabular}[c]{@{}l@{}}\rotatebox[origin=c]{90}{\makecell{SMBHyb\\\_loc}}\end{tabular}} &
\textbf{\begin{tabular}[c]{@{}l@{}}\rotatebox[origin=c]{90}{\makecell{SMBHyb\\\_rem}}\end{tabular}} &
\textbf{\begin{tabular}[c]{@{}l@{}}\rotatebox[origin=c]{90}{\makecell{SMBPor\\\_loc}}\end{tabular}} &
\textbf{\begin{tabular}[c]{@{}l@{}}\rotatebox[origin=c]{90}{\makecell{SMBPor\\\_rem}}\end{tabular}} \\ \hline
\multicolumn{1}{|l|}{\textbf{0. Attach request}} &
{\color[HTML]{C00000} Uplink} &
0 &
0 &
0 &
0 &
0 &
0 &
0 &
0 &
0 &
0 &
0 &
0 &
0 \\ \hline
\multicolumn{1}{|l|}{\textbf{1. Identity request}} &
{\color[HTML]{000000} Downlink} &
1$\pm$ 0 &
1$\pm$ 0 &
1 $\pm$ 0 &
\begin{tabular}[c]{@{}l@{}}0.9$\pm$\\ 0.3\end{tabular} &
1$\pm$ 0 &
1$\pm$ 0 &
1$\pm$ 0 &
1$\pm$ 0 &
1$\pm$ 0 &
1 $\pm$ 0 &
\begin{tabular}[c]{@{}l@{}}0.9 $\pm$ \\ 0.2\end{tabular} &
/ &
/ \\ \hline
\multicolumn{1}{|l|}{\textbf{2. Identity response}} &
{\color[HTML]{C00000} Uplink} &
31$\pm$0 &
27$\pm$6 &
\begin{tabular}[c]{@{}l@{}}38.3 $\pm$ \\ 2.3\end{tabular} &
\begin{tabular}[c]{@{}l@{}}31.0$\pm$\\ 10.4\end{tabular} &
31$\pm$0 &
\begin{tabular}[c]{@{}l@{}}31.8$\pm$\\ 2.4\end{tabular} &
\begin{tabular}[c]{@{}l@{}}25.0$\pm$\\ 6.4\end{tabular} &
31$\pm$ 0 &
31$\pm$ 0 &
\begin{tabular}[c]{@{}l@{}}31.8 $\pm$ \\ 3.5\end{tabular} &
\begin{tabular}[c]{@{}l@{}}31.0  $\pm$ \\ 4.3\end{tabular} &
/ &
/ \\ \hline
\multicolumn{1}{|l|}{\textbf{\begin{tabular}[c]{@{}l@{}}3. Authentication\\  request\end{tabular}}} &
Downlink &
1$\pm$0 &
1 $\pm$ 0 &
\begin{tabular}[c]{@{}l@{}}1.0 $\pm$ \\ 0.3\end{tabular} &
1$\pm$0 &
1$\pm$0 &
1$\pm$ 0 &
1$\pm$ 0 &
1$\pm$ 0 &
1 $\pm$ 0 &
1 $\pm$ 0 &
\begin{tabular}[c]{@{}l@{}}0.9 $\pm$\\  0.3\end{tabular} &
\begin{tabular}[c]{@{}l@{}}0.9$\pm$\\  0.1\end{tabular} &
1  $\pm$ 0 \\ \hline
\multicolumn{1}{|l|}{\textbf{\begin{tabular}[c]{@{}l@{}}4. Authentication\\  response\end{tabular}}} &
{\color[HTML]{C00000} Uplink} &
\begin{tabular}[c]{@{}l@{}}57.6 $\pm$\\ 11.4\end{tabular} &
\begin{tabular}[c]{@{}l@{}}74.1 $\pm$\\ 22.1\end{tabular} &
\begin{tabular}[c]{@{}l@{}}84.5 $\pm$ \\ 36.5\end{tabular} &
\begin{tabular}[c]{@{}l@{}}67.9$\pm$\\ 12.2\end{tabular} &
\begin{tabular}[c]{@{}l@{}}70.2$\pm$\\ 18.2\end{tabular} &
\begin{tabular}[c]{@{}l@{}}69.8 $\pm$\\ 10.0\end{tabular} &
\begin{tabular}[c]{@{}l@{}}69.1$\pm$\\ 5.9\end{tabular} &
\begin{tabular}[c]{@{}l@{}}69.9$\pm$\\ 8.2\end{tabular} &
\begin{tabular}[c]{@{}l@{}}67.9$\pm$\\ 16.2\end{tabular} &
\begin{tabular}[c]{@{}l@{}}71.7 $\pm$ \\ 10.8\end{tabular} &
{\color[HTML]{C00000} \textbf{\begin{tabular}[c]{@{}l@{}}2122.7$\pm$ \\ 309.9\end{tabular}}} &
\begin{tabular}[c]{@{}l@{}}71.2$\pm$\\ 10.7\end{tabular} &
{\color[HTML]{C00000} \textbf{\begin{tabular}[c]{@{}l@{}}1640.2$\pm$\\ 286.7\end{tabular}}}\\ \hline
\multicolumn{1}{|l|}{\textbf{\begin{tabular}[c]{@{}l@{}}5. Security mode\\  command\end{tabular}}} &
Downlink &
1$\pm$0 &
1$\pm$ 0 &
1$\pm$0 &
1$\pm$0.1 &
1$\pm$0.1 &
1$\pm$ 0 &
1$\pm$ 0 &
1$\pm$ 0 &
1 $\pm$ 0 &
1 $\pm$ 0 &
\begin{tabular}[c]{@{}l@{}}0.9 $\pm$ \\ 0.3\end{tabular} &
1  $\pm$ 0 &
1 $\pm$ 0 \\ \hline
\multicolumn{1}{|l|}{\textbf{\begin{tabular}[c]{@{}l@{}}6. Security mode\\  complete\end{tabular}}} &
{\color[HTML]{C00000} Uplink} &
\begin{tabular}[c]{@{}l@{}}20.5$\pm$\\ 3.2\end{tabular} &
\begin{tabular}[c]{@{}l@{}}19.3$\pm$ \\ 1.6\end{tabular} &
\begin{tabular}[c]{@{}l@{}}37.0$\pm$ \\ 6.3\end{tabular} &
\begin{tabular}[c]{@{}l@{}}33.0$\pm$\\ 9.5\end{tabular} &
\begin{tabular}[c]{@{}l@{}}21.8$\pm$\\ 14.3\end{tabular} &
\begin{tabular}[c]{@{}l@{}}31.3$\pm$\\ 12.5\end{tabular} &
\begin{tabular}[c]{@{}l@{}}19.6$\pm$\\ 2.6\end{tabular} &
\begin{tabular}[c]{@{}l@{}}21.9$\pm$\\ 4.6\end{tabular} &
\begin{tabular}[c]{@{}l@{}}21.8$\pm$\\ 10.5\end{tabular} &
\begin{tabular}[c]{@{}l@{}}22.4 $\pm$ \\ 5.9\end{tabular} &
\begin{tabular}[c]{@{}l@{}}20.1$\pm$ \\ 3.7\end{tabular} &
\begin{tabular}[c]{@{}l@{}}19 .7 $\pm$\\  2.7\end{tabular} &
\begin{tabular}[c]{@{}l@{}}21.1$\pm$\\ 5.8\end{tabular} \\ \hline
\multicolumn{1}{|l|}{\textbf{\begin{tabular}[c]{@{}l@{}}7. ESM information\\  request\end{tabular}}} &
Downlink &
1$\pm$0 &
1 $\pm$ 0 &
\begin{tabular}[c]{@{}l@{}}0.9 $\pm$\\ 0.2\end{tabular} &
/ &
1 $\pm$0 &
1$\pm$ 0 &
1$\pm$ 0 &
1$\pm$0 &
\begin{tabular}[c]{@{}l@{}}0.9$\pm$\\ 0.1\end{tabular} &
1 $\pm$ 0 &
1.0$\pm$ 0 &
/ &
/ \\ \hline
\multicolumn{1}{|l|}{\textbf{\begin{tabular}[c]{@{}l@{}}8. ESM information\\  response\end{tabular}}} &
{\color[HTML]{C00000} Uplink} &
19$\pm$0 &
\begin{tabular}[c]{@{}l@{}}19.7 $\pm$\\ 2.6\end{tabular} &
\begin{tabular}[c]{@{}l@{}}37.3 $\pm$ \\ 5.5\end{tabular} &
/ &
\begin{tabular}[c]{@{}l@{}}22.8$\pm$\\ 21.4\end{tabular} &
\begin{tabular}[c]{@{}l@{}}26.2$\pm$ \\ 9.3\end{tabular} &
\begin{tabular}[c]{@{}l@{}}19.6$\pm$\\ 2.3\end{tabular} &
\begin{tabular}[c]{@{}l@{}}22.6$\pm$\\ 5.6\end{tabular} &
\begin{tabular}[c]{@{}l@{}}20.7$\pm$ \\ 4.2\end{tabular} &
\begin{tabular}[c]{@{}l@{}}22.9 $\pm$ \\ 5.8\end{tabular} &
\begin{tabular}[c]{@{}l@{}}20.6 $\pm$\\ 3.9\end{tabular} &
/ &
/ \\ \hline
\multicolumn{1}{|l|}{\textbf{9. Attach accept}} &
Downlink &
\begin{tabular}[c]{@{}l@{}}50.4 $\pm$\\ 4.8\end{tabular} &
\begin{tabular}[c]{@{}l@{}}48.7$\pm$\\ 2.5\end{tabular} &
\begin{tabular}[c]{@{}l@{}}66.2 $\pm$\\ 6.8\end{tabular} &
\begin{tabular}[c]{@{}l@{}}56.9 $\pm$\\ 8.7\end{tabular} &
\begin{tabular}[c]{@{}l@{}}50.0$\pm$\\ 4.4\end{tabular} &
\begin{tabular}[c]{@{}l@{}}66.5 $\pm$\\ 14.3\end{tabular} &
\begin{tabular}[c]{@{}l@{}}50.9$\pm$\\ 5.9\end{tabular} &
\begin{tabular}[c]{@{}l@{}}48.8$\pm$\\ 3.9\end{tabular} &
\begin{tabular}[c]{@{}l@{}}49.3$\pm$ \\ 4.3\end{tabular} &
\begin{tabular}[c]{@{}l@{}}46.9 $\pm$ \\ 10.3\end{tabular} &
\begin{tabular}[c]{@{}l@{}}43.7$\pm$\\ 9.3\end{tabular} &
\begin{tabular}[c]{@{}l@{}}50.7$\pm$ \\ 6.8\end{tabular} &
\begin{tabular}[c]{@{}l@{}}57.9 $\pm$ \\ 26.1\end{tabular} \\ \hline
\multicolumn{1}{|l|}{\textbf{10. Attach complete}} &
{\color[HTML]{C00000} Uplink} &
\begin{tabular}[c]{@{}l@{}}32.4$\pm$\\ 1.9\end{tabular} &
\begin{tabular}[c]{@{}l@{}}32.8$\pm$ \\ 3.4\end{tabular} &
\begin{tabular}[c]{@{}l@{}}49.7 $\pm$ \\ 7.1\end{tabular} &
\begin{tabular}[c]{@{}l@{}}60.1$\pm$\\ 1.1\end{tabular} &
\begin{tabular}[c]{@{}l@{}}34.3$\pm$\\ 6.0\end{tabular} &
\begin{tabular}[c]{@{}l@{}}35.5 $\pm$\\ 8.2\end{tabular} &
\begin{tabular}[c]{@{}l@{}}54.5$\pm$\\ 6.4\end{tabular} &
\begin{tabular}[c]{@{}l@{}}38.8$\pm$ \\ 3.9\end{tabular} &
\begin{tabular}[c]{@{}l@{}}33.5$\pm$\\ 4.1\end{tabular} &
\begin{tabular}[c]{@{}l@{}}57.3 $\pm$ \\ 10.1\end{tabular} &
53.2 $\pm$ 9.5 &
\begin{tabular}[c]{@{}l@{}}78.5$\pm$ \\ 6.8\end{tabular} &
\begin{tabular}[c]{@{}l@{}}52.2$\pm$ \\ 4.7\end{tabular} \\ \hline
\multicolumn{2}{|l|}{\textbf{Total}} &
\begin{tabular}[c]{@{}l@{}}215.0$\pm$\\ 21.3\end{tabular} &
\begin{tabular}[c]{@{}l@{}}225.6$\pm$\\ 38.2\end{tabular} &
\begin{tabular}[c]{@{}l@{}}316.8$\pm$\\ 65.5\end{tabular} &
\begin{tabular}[c]{@{}l@{}}251.9$\pm$\\ 42.4\end{tabular} &
\begin{tabular}[c]{@{}l@{}}234.2$\pm$\\ 64.6\end{tabular} &
\begin{tabular}[c]{@{}l@{}}265.1$\pm$\\ 56.9\end{tabular} &
\begin{tabular}[c]{@{}l@{}}242.8$\pm$\\ 29.5\end{tabular} &
\begin{tabular}[c]{@{}l@{}}237.1$\pm$\\ 33.5\end{tabular} &
\begin{tabular}[c]{@{}l@{}}228.2$\pm$\\ 38.5\end{tabular} &
\begin{tabular}[c]{@{}l@{}}257.1$\pm$\\ 42.7\end{tabular} &
{\color[HTML]{C00000} \textbf{\begin{tabular}[c]{@{}l@{}}2295.5$\pm$\\ 341.7\end{tabular}}} &
\begin{tabular}[c]{@{}l@{}}253.9$\pm$\\ 31.3\end{tabular} &
{\color[HTML]{C00000} \textbf{\begin{tabular}[c]{@{}l@{}}1773.5$\pm$\\ 323.3\end{tabular}}} \\ \hline
\end{tabular}
\vspace{-1cm}
}
\end{table*}

\subsection{Decoupled devices' authentication latency}
\label{subsec:explainability}

This section investigates the latency introduced by \simbox-decoupled devices during authentication (i.e., Table \ref{tab:detailed_latency}, step 4) to uncover the cause and consistency of this overhead due to \textit{remote SIM card association}. 
To this end, we capture ME-to-SIM interactions during authentication for both a coupled device, serving as a baseline reflecting 3GPP standards, and a \simbox-decoupled device. Then by comparing these interactions, we explain decoupled devices' authentication latency, classifying its sources as either (i) specific to \simbox implementation  or (ii) imposed by standards and protocols.

\subsubsection{Methodology}

First, we describe the experimental process for capturing ME-to-SIM interactions during authentication for both a coupled device and a \simbox-decoupled device.

\vspace{0.13cm}
\noindent\textbf{Coupled devices.} Aiming to capture standardized \acrshort{ME}-to-SIM interactions during the authentication phase, we separate a coupled device's SIM card and radio processing, similarly to \textit{remote SIM card association}: As depicted in Fig. \ref{fig:methodology}, step 3 we set up a coupled network device combining
(i) a srsUE softphone, i.e., a 4G phone implemented entirely in software, running on a Linux-system PC and connecting to the shielded LTE network, (ii) a  physical SIM card within a SIM card reader connected to the softphone through an USB interface, and (iii) physical cellular antennas handled by a connected software-defined radio system (i.e., \acrshort{USRP} B210). We then perform the network attachment of the formed device and align two sets of resulting timestamped logs: (i) signaling logs at the base station and (ii) SIM card logs at the softphone.
\textit{The use of the srsUE softphone, developed in the widely adopted srsRAN 4G framework~\cite{srs4gdoc}, in our experiments attests to its generality and fidelity to 4G/LTE standards. Hence, our experiments thus provide insights into the actual implementation of 3GPP standards for the authentication phase (i.e., the \acrshort{AKA} procedure~\cite{3gpp_aka}), publicly available in \cite{srsUE_code}.}

\vspace{0.13cm}
\noindent\textbf{\simbox-decoupled devices.} 
Such devices disconnect the \acrshort{ME} (i.e., the \simbox gateway) from the SIM card (within the SIMBank), causing ME-to-SIM interactions to occur as packet exchanges over an IP network (cf. Fig \ref{fig:sig_latency}).
In order to capture these packet exchanges, we perform the network attachment of a \simbox-decoupled device and monitor packets at the control server-hosting PC using Wireshark. This setup enables us to gain insights into the interactions between the SIMBank and the gateway, which are proprietary \simbox appliances and thus typically concealed. Transport protocols (i.e., TCP or UDP) ensure reliability, order, and flow control in these IP-based interactions. We noted variations in the number of packets exchanged depending on the transport protocol used. By correlating the timing of these packets with network attachment signaling logs collected at the base station, we make specific observations for each transport protocol.

\subsubsection{Observations}
\label{subsubsec:fraud_device_obs}
From the previous experiments we make the following observations, summarized in Table \ref{tab:sim_me_interactions} in the appendix.
\begin{itemize}[leftmargin=*]
    \item First, confirming Table \ref{tab:detailed_latency} insights, authentication is the primary phase involving ME-to-SIM interactions, making it the best context for identifying any latency overhead. For coupled devices, these interactions occur \textit{only during authentication}, while \simbox-decoupled devices also show \textit{minor exchanges during the attach complete} phase. Specifically, TCP interactions involve 63 packets during authentication and 4 packets during the attach complete phase, while UDP interactions consist of 36 packets for authentication and 2 packets for attach complete (cf. Figs \ref{fig:hyb_tcp_interactions}, \ref{fig:hyb_udp_interactions}). 
    
    Shared by coupled and \simbox-decoupled devices, interactions during the authentication split into ME-to-SIM \textit{transfer} and \textit{processing} phases. Transfers consist of \textit{information transmission} from/to the ME/SIM card, while the processing phases are \textit{internal computations} within the ME/SIM card following these transfers.
    
    \item \textit{Transfers}: Logs from coupled devices reveal two physical layer round-trip transfers (four transfers in total) between the \acrshort{ME} (i.e., softphone) and the SIM card, following the ISO/IEC 7816-4 protocol~\cite{3GPP_SIM_Card} illustrated on Fig. \ref{fig:UE_internals}. These transfers are rapid, averaging 0.12 ms due to physical layer communication via \acrshort{UART} serial interfaces.\\
    In contrast, \simbox-decoupled devices show a significantly higher number of transfers. 
    They are observable with TCP while UDP's unordered nature makes them less distinguishable. Fig. \ref{fig:HYP_TCP_time_interactions} illustrates 15 transfer sessions during authentication, each involving four packets (in total 60 packets) exchanged between the \acrshort{ME} (i.e., gateway) and the SIM card in the SIMBank through the control server, which acknowledges and re-transmits the packets. On a local network, these transfers take on average 4.7 ms, totaling 70.6 ms, which 
    is a lower bound compared to real-world \simbox deployments that are Internet-based.
    
    \textit{This comparison highlights that transfer latency is a consistent indicator for distinguishing \simbox-decoupled devices.} Specifically, the 4 transfers mandated by standards result in a latency at least 39$\times$ higher than that of coupled devices, and \textit{hardly controllable due to its dependence on (i) the transport protocol and (ii) Internet vagaries. Regarding the transport protocol, while TCP increases the number of packet exchanges, UDP poorly handles network congestion, leading to retransmissions and delays.} For instance, a comparison of the latency distribution over 50 authentications of \textit{SMBHyb\_rem} configured with TCP and UDP, as shown in Fig. \ref{fig:tcp_udp}, indicates that latency with UDP is significantly higher than with TCP. Additionally, \textit{internet vagaries further contribute to transfer latency overhead.} We estimate this overhead by performing network attachment with the control server online, showing an average additional latency of 460 ms, and by measuring the median RTT of internet communication within the same country based on an empirical distribution of 1000 RTTs (cf. Fig. \ref{fig:rtt_latency}). The median value of 57.4 ms suggests that\textit{ the 2 RTTs imposed by the standard guarantee a transfer latency overhead of 114.8 ms for \simbox-decoupled devices, which is almost twice the avg. auth. latency of coupled devices (cf. Table \ref{tab:detailed_latency}).}
    \begin{figure*}
    \centering
    \includegraphics[width=\linewidth]{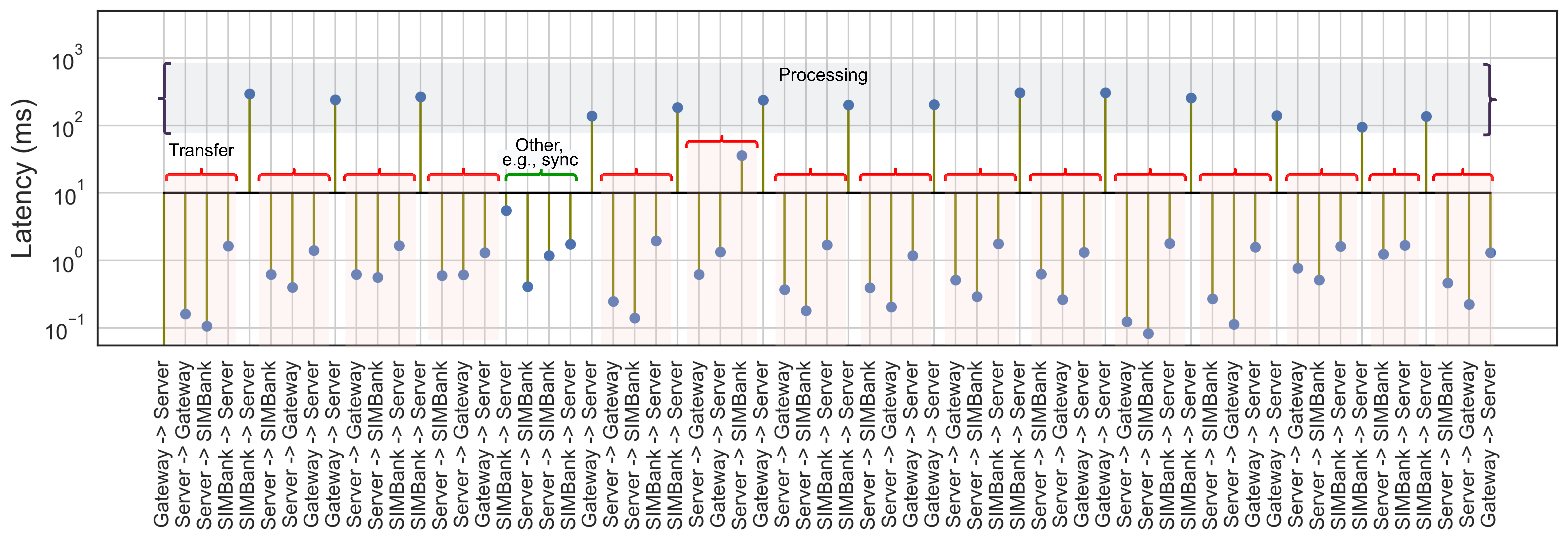}
    \caption{Hybertone \simbox components TCP interactions during the authentication phase.}
    \label{fig:HYP_TCP_time_interactions}
    \vspace{-0.3cm}
    \end{figure*}

    \item \textit{Processing}-induced latency is much higher than transfer latency.
    Analysis of coupled devices' logs reveals two standard-imposed processing phases on the SIM card and three on the \acrshort{ME}, averaging 15.6 ms and 9.4 ms respectively (cf. Table \ref{tab:sim_me_interactions}).
    In contrast, \simbox-decoupled devices exhibit as many as 14 processing phases (8 for the SIM card and 6 for the \acrshort{ME}), averaging 218 ms and 211 ms respectively with TCP, and 12 (6 each for the SIM card and the \acrshort{ME}), averaging 236.1 ms and 139.3 ms respectively with UDP.
     
    \textit{This comparison highlights that the elevated processing latency in \simbox-decoupled devices is likely due to their implementation involving a higher number of processing phases than necessary and significantly longer average times. Although this could be optimized by \simbox manufacturers, some overhead may be inevitable due to the simultaneous control of multiple SIM cards and GSM modules, as well as the encapsulation/decapsulation of information exchanged during authentication into IP packets. This overhead is challenging to quantify for proprietary devices.}
\end{itemize}

\greybox{\noindent\textbf{Insight.} \textit{In essence, our investigations show that \simbox-decoupled devices' authentication latency is influenced by factors beyond \simbox owners' control. Even with optimization efforts, i.e., reducing the transfer count and processing time, this latency cannot match that of stripped-down, coupled devices, maintaining a consistent distinction between \simbox-decoupled devices and their coupled counterparts.} }

\subsection{Coupled devices' authentication latency}
\label{subsec:tresholding}

Here, we assess the feasibility of instances where a coupled device's latency could be high enough to be mistaken for a \simbox-decoupled one. To this end, we scrutinize the latency of coupled devices during the authentication phase by breaking down a coupled device's auth. latency into two components:
(i) \textit{A transmission latency} that includes the wireless propagation time along with any delay related to the wireless network channel condition (\S \ref{subsubsec:transmission}) and 
(ii) \textit{A processing latency} in which the device locally runs the authentication algorithm until a response is generated (\S \ref{subsubsec:processing}). Note that these latencies differ from the ones detailed in \S \ref{subsubsec:fraud_device_obs}, which focused on intra-device interactions. The current context instead aims to analyze the latency in the device's communication with the network through the base station.


\subsubsection{Transmission latency of the authentication}
\label{subsubsec:transmission}

Transmission latency refers to the back-and-forth communication time between the device and the base station during an authentication request and response. 
Predicting this latency is challenging due to various factors that affect the quality of each device's experience. In particular, although the \textit{wireless signal propagation} time is negligible as the signal moves at the speed of light, and LTE employs an admission control mechanism~\cite{AdmissionControl} to prevent delays during the attachment caused by \textit{network congestion}, the impact of \textit{signal quality} on authentication latency remains to be determined.

Indeed, poor signal quality, indicated by an LTE \acrshort{RSRP} of less than -110 dBm, often results from significant distance to the cell tower, interference, or device sensitivity issues. This can cause re-transmissions and signaling delays. In the following, we assess this impact on authentication latency by measuring it in an outdoor network and generalizing the results in a controlled indoor testbed (cf. Fig. \ref{fig:methodology}, step 4).
\vspace{-0.13cm}
\paragraph{\textbf{Methodology.}} 
Regarding \textit{outdoor signal attenuation}, we measure the outdoor signal quality over three days with varying weather conditions (sunny, rainy, and windy) using a Samsung Galaxy Note4 (referred to as \textit{GalaxyNote4}) in a vehicle covering over 80km of city roads in a central urban area in the Paris region. As open phone signaling data is not public, we use the QCSuper open-source diagnostic logging tool~\cite{QCSuper}, which is compatible only with Qualcomm-based phones, to capture the network attachment signaling messages. 
Data collection is done with a QCSuper-installed laptop connected to a \textit{GalaxyNote4} phone, the only compatible device in our testbed. 
The collected outdoor dataset comprises 2287 network attachments of the \textit{GalaxyNote4} phone. Each network attachment marks the phone's entrance into a new network cell 
and induces an authentication process. From the collected logs, we extract signal quality measurements (i.e., \acrshort{RSRP}) within the respective cell for each network attachment.

To extend these findings across various phone models, we replicate the network \textit{indoor signal attenuation} in a controlled environment (cf. \S \ref{subsec:experimental_setup}) using static attenuators.
Specifically, we apply two attenuation levels to the initial network signal quality of -71 dBm (excellent quality), resulting in measurements of (i) -90 dBm (medium quality) and (ii) -100 dBm (poor quality) (cf. Fig. \ref{fig:spectrum}, in the appendix). 
For each signal quality condition, we conduct 50 network attachments for every phone model (described in Table \ref{tab:testbed}), including the \textit{GalaxyNote4} phone used in the outdoor measurement scenario, and record the corresponding authentication latency.

\vspace{-1em}
\paragraph{\textbf{Observations.}}  
In Fig. \ref{fig:outdoor_latency_steps}, we break down the attachment latency distribution for the \textit{GalaxyNote4} in the outdoor scenario, detailing each step of the network attachment process. Since latency data is collected at the device level, our focus is on the downlink messages (UE←BS), which include the transmission latency of interest. Specifically, we examine the impact of signal quality (i.e., \acrshort{RSRP}) on the latency of the \textit{"security mode command"} message, which inherently captures the transmission latency of the \textit{"authentication response."}  To streamline interpretation, we approximate the \textit{"security mode command"} latency as the transmission latency of the authentication process in Fig. \ref{fig:outdoor_rsrp_security}. With signal quality measurements ranging from -65 to -119 dBm, our results cover all radio frequency conditions, from "Cell Edge" to "Excellent"~\cite{3GPP_RSRP_range}, ensuring the representativeness and depth of our findings.

Notably, Fig. \ref{fig:outdoor_rsrp_security} shows that the upper limit for the auth. response's transmission latency is negligible. Although some outliers around 200 ms, the majority of the values indicate low latency, irrespective of meteorological conditions (denoted by the days) or signal quality (denoted by the \acrshort{RSRP}). \textit{This result convincingly shows that signal quality has a minimal impact on authentication transmission latency. Furthermore, linear regression of latency and signal quality confirms this trend across a broader signal quality spectrum.} 

Figure \ref{fig:attenuation_cage}, shows the authentication latency under indoor signal attenuation, extending our findings to other phone models. The results indicate that signal quality has a negligible impact on the authentication latency for the \textit{GalaxyNote4, GalaxyS3,} and \textit{GalaxyZFold5G}, confirming our earlier interpretations from outdoor scenarios. However, the six remaining phone models did not initiate network attachment under medium to high signal attenuation, likely due to the lower sensitivity of their receivers. In a carrier network, these phones would have connected to nearby cells with stronger signals, thereby avoiding any latency issues related to signal quality.

\vspace{0.13cm}
\greybox{\noindent \textbf{Insight:} \textit{
Our findings demonstrate that the round-trip transmission latency during the authentication phase is negligible and robust, showing little sensitivity to fluctuations in network signal quality. Even in the worst-case scenario, poor signal quality will cause mobile devices to switch network cells rather than increase transmission latency.}}

\begin{table*}
\begin{minipage}{0.24\linewidth}
\centering
\includegraphics[width=\linewidth]{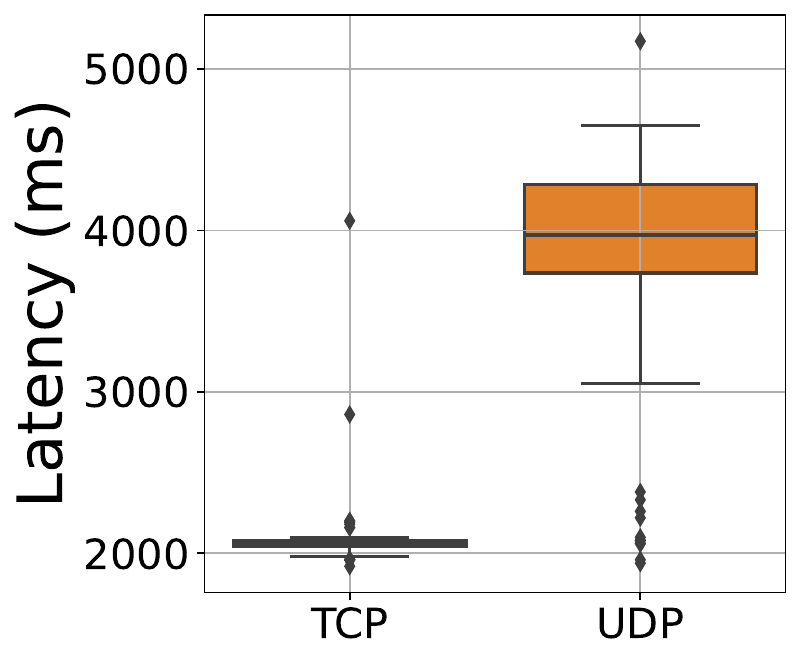}
\captionof{figure}{\textit{SMBHYB\_rem} TCP vs UDP auth. latency.}
\label{fig:tcp_udp}
\end{minipage}
\hfill
\begin{minipage}{0.25\linewidth}
\centering
\includegraphics[width=0.95\linewidth]{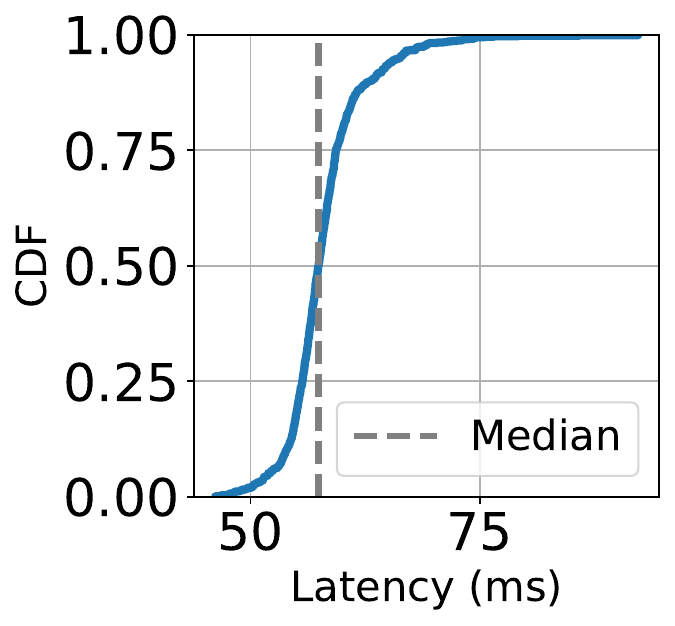}
\captionof{figure}{RTT latency distribution over Internet}
\label{fig:rtt_latency}
\end{minipage}
\hfill
\begin{minipage}{0.48\linewidth}
\centering
\includegraphics[scale=0.45]{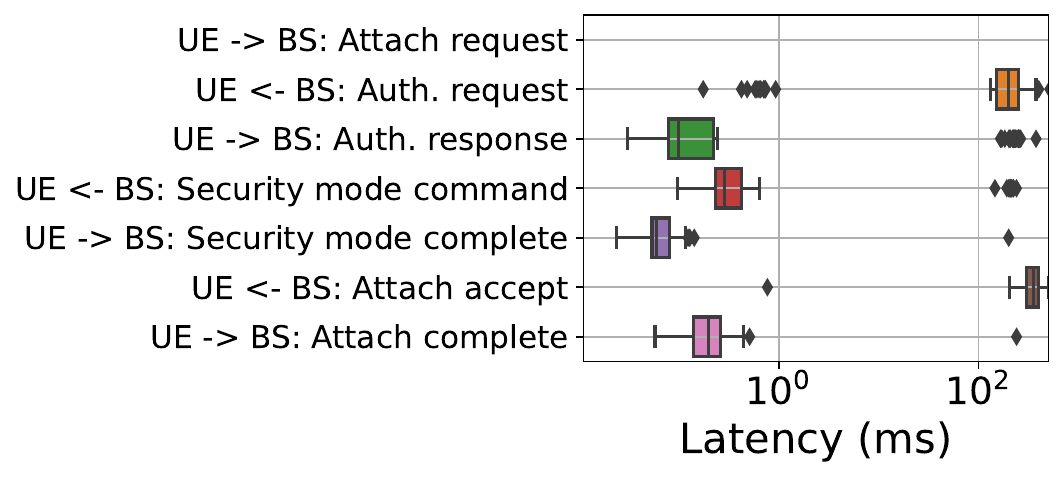}
\captionof{figure}{Latency (in log-scale) in different attachment steps for \textit{GalaxyNote4}'s outdoor scenario}
\label{fig:outdoor_latency_steps}
\end{minipage}
\end{table*}
\begin{table*}
\begin{minipage}{0.32\linewidth}
\includegraphics[width=0.98\linewidth]{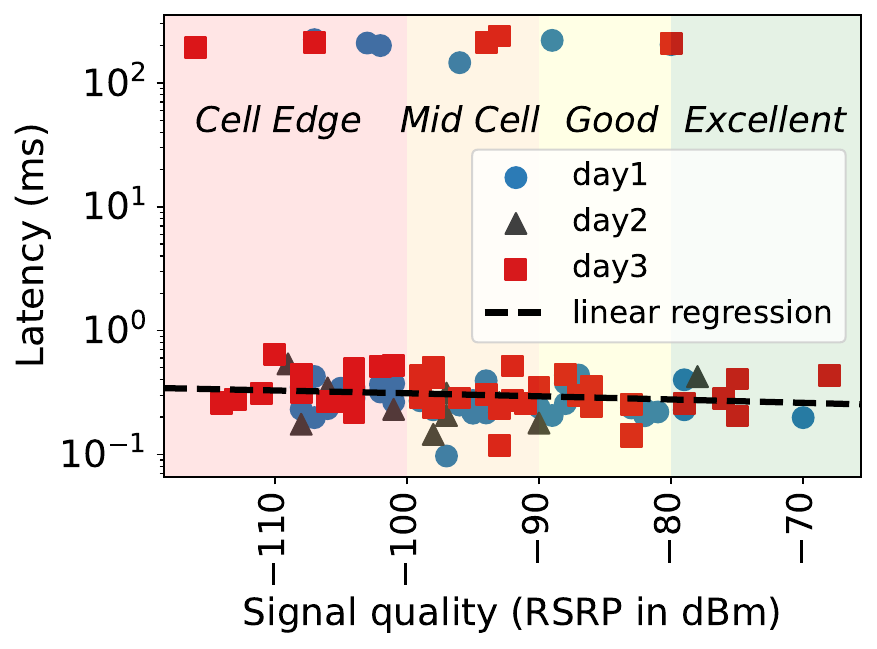}
\captionof{figure}{Outdoor transmission latency (in log-scale) of the authentication response w.r.t. the signal quality.}
\label{fig:outdoor_rsrp_security}
\end{minipage}
\hfill
\begin{minipage}{0.4\linewidth}
\centering
\includegraphics[width=0.94\linewidth]{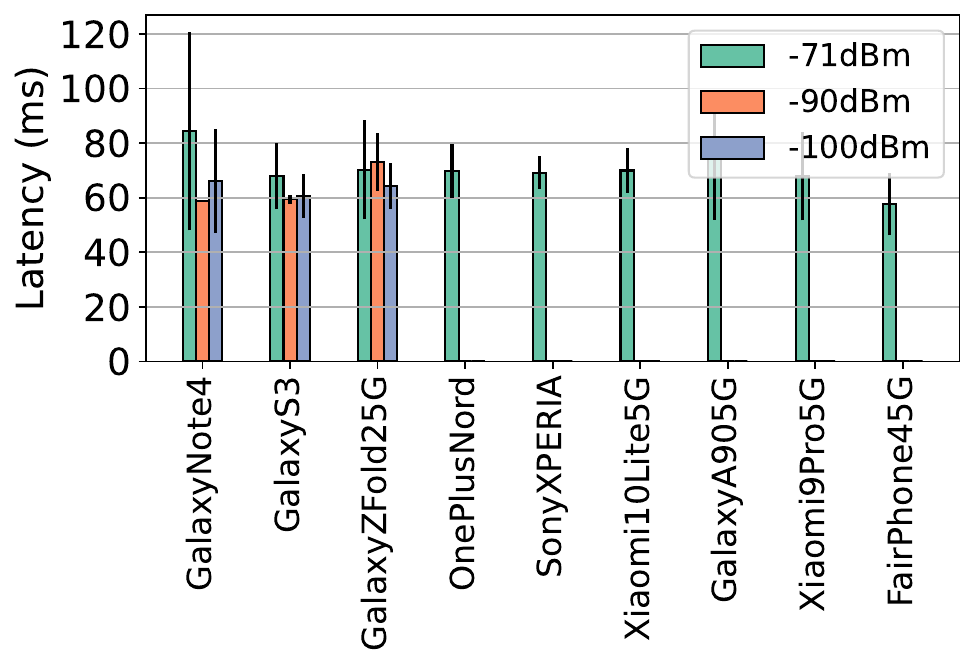}
\captionof{figure}{Indoor network authentication latency per phone model w.r.t. the signal quality.}
\label{fig:attenuation_cage}
\end{minipage}
\hfill
\begin{minipage}{0.22\linewidth}
\centering
\includegraphics[width=\linewidth]{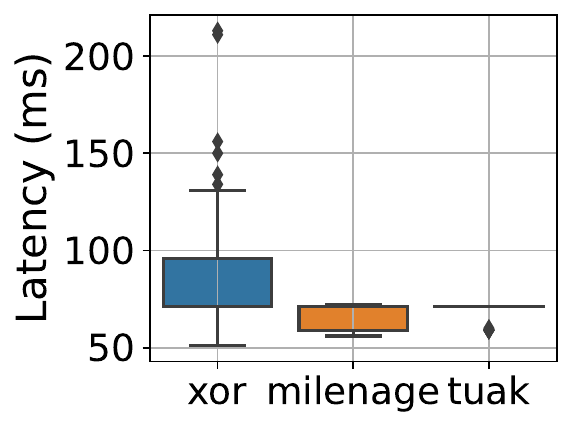}
\captionof{figure}{\textit{GalaxyNote4}'s authentication latency variation w.r.t. the SIM authentication algorithm.}
\label{fig:auth_algo}
\end{minipage}
\end{table*}

\subsubsection{Processing latency of the authentication}
\label{subsubsec:processing}

The processing latency denotes the time required for authentication computations within coupled devices, specifically within the SIM cards provided by the operator. As a result, it barely varies across different phone models or depends on phone features (i.e., processor, battery, or RAM), since modern phones are designed to handle much more resource-intensive applications.
Thus, the authentication processing latency is primarily determined by the \textit{SIM card  dauthentication algorithm}, which runs inside the SIM card to compute the expected network Authentication Response (RES). Chosen by each operator and kept secret to prevent account impersonation, these algorithms are typically variants of standardized algorithms like XOR~\cite{3GPP_Xor}, Milenage~\cite{ETSI_milenage}, or Tuak~\cite{3GPP_Tuak}.

We evaluate the impact of the standardized SIM card authentication algorithms on the authentication latency. Specifically, we configure inside our indoor testbed (cf. \S \ref{subsec:experimental_setup}) such different authentication algorithms on the \textit{GalaxyNote4} phone and conduct 50 network attachments for each algorithm, recording the resulting authentication latency values. The findings in Fig. \ref{fig:auth_algo} confirm that the authentication algorithm impacts both the processing latency and its variability. \textit{Thus, mobile operators can achieve lower processing latency by carefully selecting their SIM authentication algorithm.}

\section{SigN implementation}
\label{sec:real_world_deployment}
This section delves into utilizing the authentication latency metric for the practical implementation of \sign for \simbox activity detection at the mobile edge. 

\vspace{0.13cm}
\noindent\textbf{Supporting insights.}
The experiments conducted in \S \ref{sec:latency_detection} underscored a \textit{significant distinction} in attachment latency between coupled and \simbox-decoupled devices, specifically \textit{during the authentication phase} (cf. Table \ref{tab:detailed_latency}). Acknowledging its variability influenced by internal/external factors, we further investigated the authentication latency of both coupled and \simbox-decoupled devices. First, \textit{coupled devices authentication latency in outdoor networks remains within a consistent range and is markedly lower than observed for \simbox-decoupled devices in indoor settings} (cf. Fig. \ref{fig:outdoor_latency_steps}).
Second, our investigations reveal that \textit{a non-negligible portion of the observed authentication latency in \simbox-decoupled devices is imposed by factors beyond fraudsters' control}, such as LTE standards and Internet-based communication protocols and vagaries (cf. \S \ref{subsec:explainability}). Moreover, like coupled devices, \textit{\simbox-decoupled devices experience additional transmission latency} in the real-life conditions of operator networks. Given these findings, \textit{\textbf{monitoring authentication latency at the network edge proves to be a reliable and practical method for distinguishing \simbox activity from regular one.}} 

\vspace{0.13cm}
\noindent\textbf{Monitoring the authentication latency.} 
In LTE, the authentication procedure initiates the monitoring of the induced latency through a logging mechanism. Consequently, the latency of each authentication is automatically logged at each base station, and is useful for network performance monitoring, optimization, and Quality of Service (QoS) management.

However, \textit{experiments in this paper highlight that authentication latency is not only an indicator of network QoS but also a robust metric for identifying vulnerabilities related to \simbox activity}. Our findings demonstrate that, while the latency is generally acceptable (i.e., within the 6s timer range~\cite{3GPP_Auth_timer}) for both coupled and \simbox-decoupled devices, it creates a clear distinction between \simbox activity and regular one. 

Therefore, \textit{\sign approach suggests a new monitoring usage of authentication latency in cellular networks, allowing the detection of \simbox activity through its distribution.} According to 3GPP standards, \textbf{network operators have the flexibility to initiate at any time an authentication procedure when a signaling connection with a device exists}~\cite{3GPP_Auth_timer}. This flexibility enables operators to capture the distribution of mobile devices' authentication latency by initiating multiple authentications at different times throughout the day. Randomly timed authentications are essential to accurately reflect a device’s behavior, as passive collection could be manipulated by \simbox operators who might switch between remote and local SIM card associations to skew the latency distribution.

Relying on the distribution rather than a single measurement is crucial for ensuring robustness against high-value outliers that may occur for coupled devices (cf. Fig. \ref{fig:outdoor_rsrp_security}). Hence, analyzing a full day’s data yields key metrics such as the mean, median, and standard deviation of authentication latencies, highlighting \textit{devices with consistently unusually high values and prompting further investigation by the operator.} 

Such monitoring is lightweight, leveraging existing automatic functions in cellular networks, i.e., logs collected by each base station. As a result, it allows operators to make informed decisions without the network edge's overhead.

\begin{figure}
    \centering
    \includegraphics[width=0.9\linewidth]{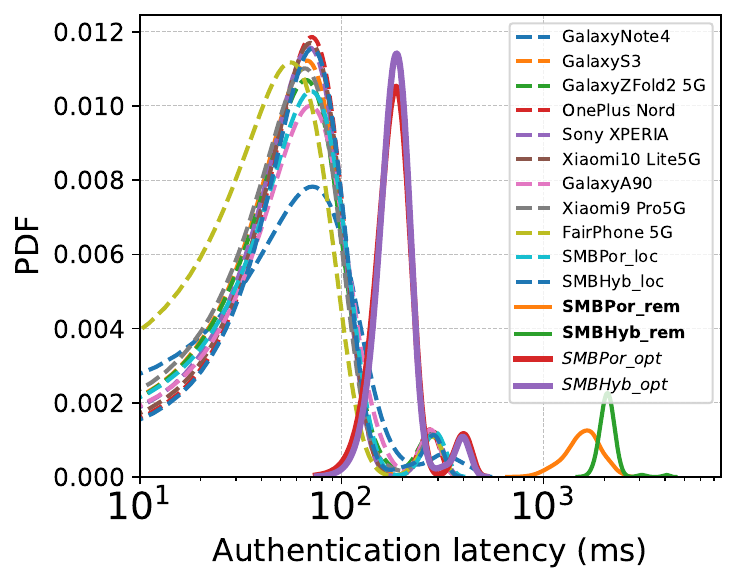}
    \caption{Distribution of authentication latency of (i) coupled devices (\textit{dashed}), (ii) current (\textit{bold label}) and (iii) optimized (\textit{italic label}) \simbox-decoupled devices }
    \label{fig:latency_pdf}
    \vspace{-0.6cm}
\end{figure}

\vspace{0.13cm}
\noindent\textbf{Statistical support.} 
We statistically validate the effectiveness of the \sign approach from our network measurements. 
Fig. \ref{fig:latency_pdf} plots realistic distributions of authentication latency for: (i) coupled devices with outdoor transmission latency, (ii) Current measurements of \simbox-decoupled devices with outdoor transmission latency, and (iii) Optimized \simbox-decoupled devices (cf. \ref{subsubsec:fraud_device_obs}) consisting of \simbox coupled device with outdoor transmission latency and simulated reduced ME-to-SIM transfers (2 RTTs as on Fig. \ref{fig:rtt_latency}).
We employed a \textit{t-test}, a key statistical tool, to compare the means of coupled and \simbox-decoupled devices (both current and optimized). Details are provided in the appendix (cf. \S \ref{sec:ttest}). 

The \textit{t-ratio} of 15.29 (with $t=25.27$ and \textit{critical value} $=1.65$) reveals a significant statistical difference between the coupled and \simbox-decoupled device groups. The \textit{p-value} of $1.3\times 10^{-102}$ indicates an almost zero chance of overlap between the two groups and  
a high probability of correctly identifying the attachment activity even of the most optimized \simbox-decoupled device. 


\section{Limitations}
\label{sec:limitations}
Acknowledging the limitations of this study is crucial to contextualizing its findings and guiding future research efforts. While we believe our approach and results provide valuable insights into monitoring \simbox activity in mobile networks, there are areas where constraints, assumptions, and specific conditions may have influenced the outcomes.

First, due to the proprietary nature of \simbox devices, we were unable to perform an in-depth reverse engineering of their hardware and software. This limitation constrained our ability to fully uncover the specific implementation techniques used by \simbox manufacturers. Instead, we adopted a standards-based approach to evaluate potential baseline measures that fraudsters must adhere to. 
While our approach sheds light on baseline security assumptions, future research could explore advanced logic analysis techniques to extract and study SIM card secrets from these devices, providing deeper insights into their operation.

Second, while our experimentation was conducted on a simulated 4G network powered by the Amarisoft suite, we acknowledge that this does not fully replicate the complexities of a live operator network. However, the testbed design and the use of Amarisoft's professional-grade software, widely trusted by Mobile Network Operators (MNOs), ensure results that closely approximate those observed in real-world conditions. Additionally, testing in a controlled environment allowed for precise measurements and analysis without the risk of interfering with live operator networks. Despite this, future validations on live networks could provide additional insights into practical deployment scenarios.

Lastly, the TCP and UDP latency measurements presented in Figure~\ref{fig:rtt_latency} are derived from a single network configuration, which may not fully capture the variability across different operator networks. While these measurements are representative and align with general expectations, they may overestimate latency in certain cases. We believe this limitation can be mitigated by adapting our detection method to each operator's specific network conditions through a straightforward calibration process. Future work should consider extending these measurements across diverse network environments to enhance the generalizability of our findings.

\vspace{-0.3cm}

\section{Conclusion}
\label{sec:conclude}
This paper introduced \sign, an online prevention solution to uncover \simbox activity at the cellular edge. Based on an empirical study of network attachment latency in coupled and \simbox-decoupled devices, we found that \simbox-decoupled devices exhibit higher authentication latency. \sign optimizes existing cellular monitoring, improving \simbox activity detection efficiency.

\sign's significance lies in its effectiveness and practicality, enabling easy integration into operator networks. This offers substantial economic benefits and resolves challenges faced by current network-edge solutions, making \sign a key advancement in securing networks against \simbox activity.

Note that while this paper measurements focus on smartphones, the findings apply to other devices with a physical or embedded SIM card, including tablets, laptops, and IoT devices. The key distinction is the separation of the SIM card from the Mobile Equipment in \simbox-decoupled devices.

\bibliographystyle{ACM-Reference-Format}
\bibliography{references}

\appendix
\newacronym{UART}{UART}{Universal Asynchronous Receiver/Transceiver} 
\newacronym{CDR}{CDR}{Call Detail Records}
\newacronym{GSM}{GSM}{Global System for Mobile Communications}
\newacronym{ESM}{ESM}{EPS Session Management}
\newacronym{EPS}{EPS}{Evolved Packet System}
\newacronym{AKA}{AKA}{Authentication and Key Agreement}
\newacronym{ME}{ME}{Mobile Equipment}
\newacronym{NAS}{NAS}{Non-Access Stratum}
\newacronym{UE}{UE}{User Equipment}
\newacronym{TAC}{TAC}{Type Allocation Code}
\newacronym{IMEI}{IMEI}{International Mobile Equipment Identity}
\newacronym{HBS}{HBS}{Human Behavior Simulation}
\newacronym{CA}{CA}{Carrier Aggregation}
\newacronym{SDR}{SDR}{Software-Defined Radio}
\newacronym{MME}{MME}{Mobility Management Entity}
\newacronym{IMS}{IMS}{IP Multimedia Subsystem}
\newacronym{SGW}{SGW}{Serving Gateway}
\newacronym{USRP}{USRP}{Universal Software Radio Peripheral}
\newacronym{RSRP}{RSRP}{Reference Signal Received Power}
\newacronym{PLMN}{PLMN}{Public Land Mobile Network}
\newacronym{AUTN}{AUTN}{Authentication Token}
\newacronym{XRES}{XRES}{Expected Response}
\newacronym{KASME}{KASME}{Key Agreement Key Stored in the \acrlong{ME}}
\newacronym{IMSI}{IMSI}{International Mobile Subscriber Identity}

\printglossary[type=\acronymtype]

\section{T-Test Procedure and Statistical Analysis}
\label{sec:ttest}



To perform a t-test, we first calculate the means of the two groups ($\mu_l$ and $\mu_f$), their respective sizes ($n_l$ and $n_f$), and the pooled standard error ($SE$) of the two groups, as shown in equation \ref{eq:se}. The t-statistic ($t$) is then computed as the ratio of the difference between the group means to the pooled standard error, as detailed in equation \ref{eq:t}.

\begin{multicols}{2}
  \begin{equation}
  \label{eq:se}
    SE = \sqrt{\left(\frac{\sigma_{l}^2}{n_{l}}\right) + \left(\frac{\sigma_{f}^2}{n_{f}}\right)}
  \end{equation}\break
  \begin{equation}
  \label{eq:t}
     t = \lvert \frac{\mu_{f} - \mu_{l}}{\sqrt{SE^2(\frac{1}{n_f}+\frac{1}{n_l})}} \rvert
  \end{equation}
\end{multicols}

\noindent Unlike the classical standard deviation ($\sigma_l$ or $\sigma_f$), which measures dispersion within each group, the pooled standard error ($SE$) incorporates variability from both groups to provide a more accurate estimate of the population standard deviation.

A higher t-statistic value suggests a more significant difference between the groups. To determine the statistical significance of this difference, we compare the computed t-statistic to a \textit{critical value}, which is typically based on a predetermined confidence level (e.g., 95\%). If the computed t-statistic exceeds this critical value, it indicates that the observed difference is statistically significant and unlikely due to random chance, thereby supporting the validity of the findings.

\begin{table}
\begin{minipage}{\linewidth}
\centering
\includegraphics[width=0.8\linewidth]{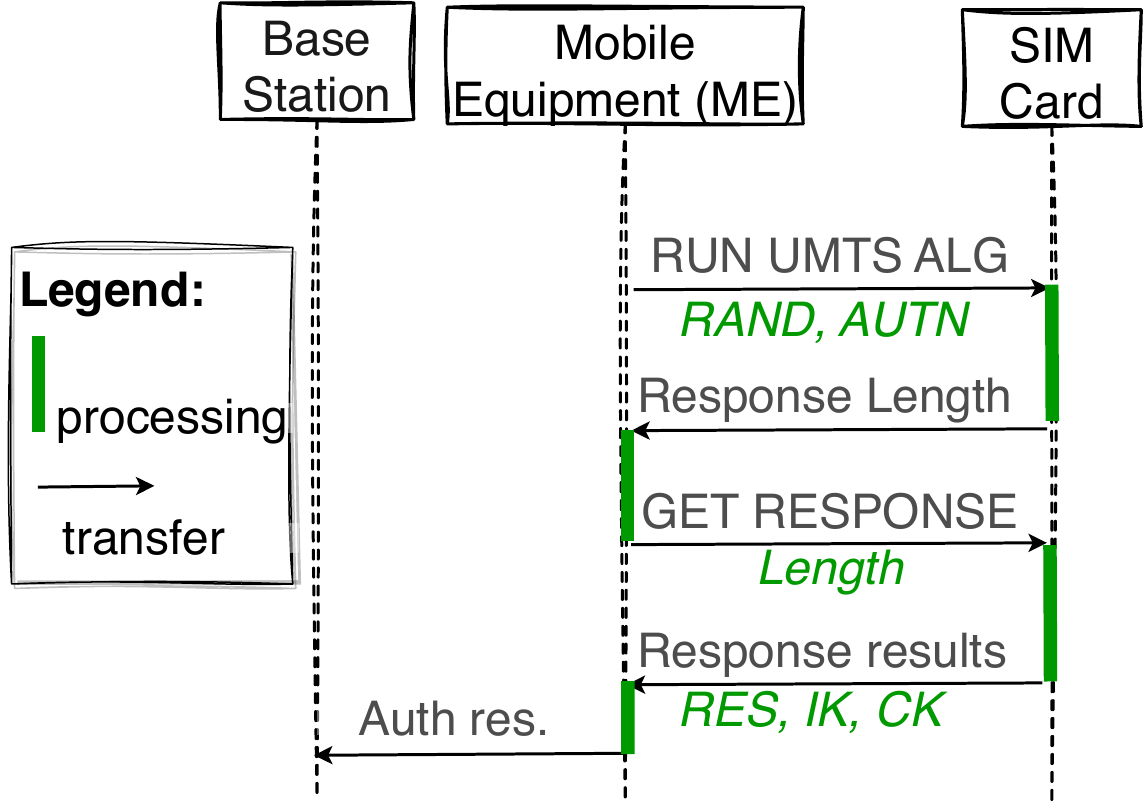}
\captionof{figure}{coupled \acrshort{ME} to SIM card interactions during the authentication}
\label{fig:UE_internals}
\end{minipage}
\end{table}

\begin{figure}
\centering
\begin{subfigure}{0.5\textwidth}
\centering
\includegraphics[scale=0.5]{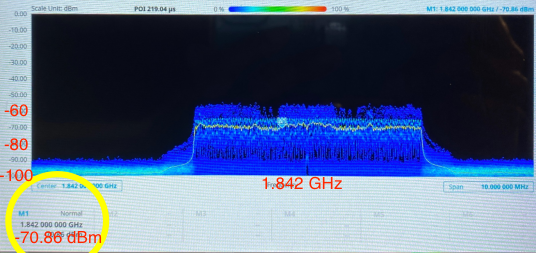}
\caption{Reference state: -71 dBm} 
\label{fig:spectrum0}
\end{subfigure}
\hfill
\begin{subfigure}{0.5\textwidth}
\centering
\includegraphics[scale=0.5]{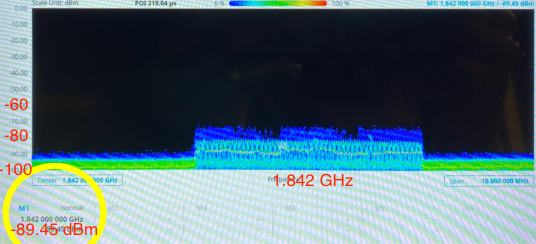}
\caption{-90dBm RSRP attenuation} 
\label{fig:spectrum1}
\end{subfigure}
\hfill
\begin{subfigure}{0.5\textwidth}
\centering
\includegraphics[scale=0.5]{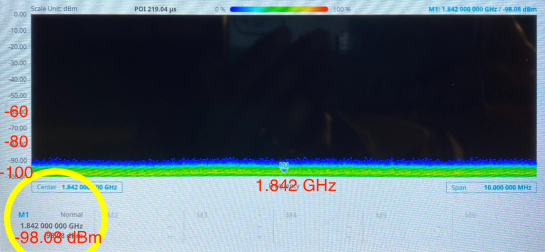}
\caption{-100dBm RSRP attenuation} 
\label{fig:spectrum2}
\end{subfigure}
\caption{\acrfull{RSRP} measurement  inside the testbed: x-axis: RSRP (dBm), y-axis: Frequency (Hz)}
\label{fig:spectrum}
\end{figure}

\begin{table*}
\centering
\caption{Testbed component specifications}
\label{tab:testbed}
\resizebox{\textwidth}{!}{%
\begin{tabular}{|ll|l|}
\hline
\multicolumn{2}{|l|}{\textbf{Parameters}}             & \textbf{Values}                                  \\ \hline
\multicolumn{2}{|l|}{\begin{tabular}[c]{@{}l@{}}Host PC (BS, MME, SGW)\end{tabular}} &
  \begin{tabular}[c]{@{}l@{}}Intel(R) Core(TM) i9-10900K CPU@3.70GHz, 16GB RAM, GB Ethernet controller\end{tabular} \\ \hline
\multicolumn{1}{|l|}{}         & Bandwidth            & 5MHz FDD                                         \\ \cline{2-3} 
\multicolumn{1}{|l|}{}         & Configuration        & SISO (Single Input Single Output)                                          \\ \cline{2-3} 
\multicolumn{1}{|l|}{\multirow{-3}{*}{Cell}} &
  Frequency &
  \begin{tabular}[c]{@{}l@{}}Downlink center frequency: 1845 MHz, Band 3\end{tabular} \\ \hline
\multicolumn{2}{|l|}{\begin{tabular}[c]{@{}l@{}}Programmable SIM cards\end{tabular}} &
  Sysmocom SysmoSIM-SJS1 \\ \hline
\multicolumn{2}{|l|}{}                                & Samsung Galaxy Note 4 (x3)                       \\ \cline{3-3} 
\multicolumn{2}{|l|}{}                                & Samsung Galaxy S3                                \\ \cline{3-3} 
\multicolumn{2}{|l|}{}                                & Xiaomi Redmi Note 9                              \\ \cline{3-3} 
\multicolumn{2}{|l|}{}                                & Xiaomi 10 Lite 5G (x2)                           \\ \cline{3-3} 
\multicolumn{2}{|l|}{}                                & FairPhone 4 5G                                   \\ \cline{3-3} 
\multicolumn{2}{|l|}{}                                & OnePlus Nord Model 5G                            \\ \cline{3-3} 
\multicolumn{2}{|l|}{}                                & Sony XPERIA                                      \\ \cline{3-3} 
\multicolumn{2}{|l|}{}                                & {\color[HTML]{333333} Samsung galaxy Z Fold2 5G} \\ \cline{3-3} 
\multicolumn{2}{|l|}{\multirow{-9}{*}{Mobile Phones}} & Samsung Galaxy A90 5G                            \\ \hline
\multicolumn{2}{|l|}{} &
  \begin{tabular}[c]{@{}l@{}}Hybertone\\ - SIMBank: SMB32 - Gateway: GoIP8 (x2)\\- Control server v. 2022-5-11 (Host PC:
  Intel(R) Core(TM) i5-4590 CPU @ 3.30GHz, 8GB RAM, GB Ethernet controller)\end{tabular} \\ \cline{3-3} 
\multicolumn{2}{|l|}{\multirow{-2}{*}{\simbox appliances}} &
  \begin{tabular}[c]{@{}l@{}}Portech\\- SIMBank: SBK-32 - Gateway: MV-374\\- Control server SS-128 (Host PC:
    Intel(R) Core(TM) i7-4610M CPU @ 3.00GHz, 16GB RAM, GB Ethernet controller)  \end{tabular} \\ \hline
    \end{tabular}%
}
\end{table*}

\begin{table*}[]
\centering
\caption{Finegrained analysis of authentication latency (in ms)}
\label{tab:sim_me_interactions}
\resizebox{\textwidth}{!}{%
\begin{tabular}{ll|cllcll|cll|cll|}
\cline{3-14}
\multicolumn{2}{l|}{} &
  \multicolumn{6}{c|}{\textit{SMBHyb\_rem}} &
  \multicolumn{3}{c|}{\textit{SMBPor\_rem}} &
  \multicolumn{3}{c|}{} \\ \cline{3-11}
\multicolumn{2}{l|}{\multirow{-2}{*}{}} &
  \multicolumn{3}{c|}{\textit{TCP}} &
  \multicolumn{3}{c|}{\textit{UDP}} &
  \multicolumn{3}{c|}{\textit{UDP}} &
  \multicolumn{3}{c|}{\multirow{-2}{*}{\textit{srsUE softphone}}} \\ \hline
\multicolumn{1}{|l|}{\textbf{Step}} &
  \textbf{Dir.} &
  \multicolumn{1}{c|}{\textbf{latency}} &
  \multicolumn{1}{c|}{\textbf{Transfer}} &
  \multicolumn{1}{c|}{\textbf{Processing}} &
  \multicolumn{1}{c|}{\textbf{latency}} &
  \multicolumn{1}{c|}{\textbf{Transfer}} &
  \multicolumn{1}{c|}{\textbf{Processing}} &
  \multicolumn{1}{c|}{\textbf{latency}} &
  \multicolumn{1}{c|}{\textbf{Transfer}} &
  \multicolumn{1}{c|}{\textbf{Processing}} &
  \multicolumn{1}{c|}{\textbf{latency}} &
  \multicolumn{1}{c|}{\textbf{Transfer}} &
  \multicolumn{1}{c|}{\textbf{Processing}} \\ \hline
\multicolumn{1}{|l|}{\textbf{\begin{tabular}[c]{@{}l@{}}4. Authen-\\tication \\ response\end{tabular}}} &
  {\color[HTML]{C00000} Uplink} &
  \multicolumn{1}{l|}{3259} &
  \multicolumn{1}{l|}{\begin{tabular}[c]{@{}l@{}}15 sessions\\ 4.7 $\pm$ 9.2\\ total: 70.6\end{tabular}} &
  \multicolumn{1}{l|}{\begin{tabular}[c]{@{}l@{}}14 occurences:\\ \\ - SIMBank (8)\\ 218$\pm$8\\ total: 1744.3\\ \\ - Gateway (6)\\ 211$\pm$6\\ total: 1265.8\end{tabular}} &
  \multicolumn{1}{l|}{2379} &
  \multicolumn{1}{l|}{\begin{tabular}[c]{@{}l@{}}Not clearly\\ identified\end{tabular}} &
  \begin{tabular}[c]{@{}l@{}}12 occurences:\\ \\ - SIMBank (6)\\ 236.1$\pm$116.3\\ total:1416.4\\ \\ - Gateway (6)\\ 139.3$\pm$73.6\\ total: 835.9\end{tabular} &
  \multicolumn{1}{l|}{1199} &
  \multicolumn{1}{l|}{\begin{tabular}[c]{@{}l@{}}1 session\\ total: 4.2\end{tabular}} &
  \begin{tabular}[c]{@{}l@{}}Not clearly\\ identified\\ - Before\\transfer\\ 774.4\\ - After\\transfer\\ 420.4\end{tabular} &
  \multicolumn{1}{l|}{59} &
  \multicolumn{1}{l|}{\begin{tabular}[c]{@{}l@{}}4 sessions\\ 0.12$\pm$0.15\\ total: 0.58\end{tabular}} &
  \begin{tabular}[c]{@{}l@{}}5 occurences\\ \\ - SIM card (2)\\15.6$\pm$14.5\\ total: 31.1\\ \\ - \acrshort{ME} (3)\\ 9.4$\pm$10.8\\ total: 28.1\end{tabular} \\ \hline
\multicolumn{1}{|l|}{\textbf{\begin{tabular}[c]{@{}l@{}}10. Attach \\ complete\end{tabular}}} &
  {\color[HTML]{C00000} Uplink} &
  \multicolumn{1}{l|}{40} &
  \multicolumn{1}{l|}{\begin{tabular}[c]{@{}l@{}}1 session\\ total: 2.8\end{tabular}} &
  \multicolumn{1}{l|}{/} &
  \multicolumn{1}{l|}{60} &
  \multicolumn{1}{l|}{\begin{tabular}[c]{@{}l@{}}1 session\\ total: 0.2\end{tabular}} &
  / &
  \multicolumn{1}{l|}{52} &
  \multicolumn{1}{l|}{/} &
  / &
  \multicolumn{1}{l|}{48} &
  \multicolumn{1}{l|}{/} &
  / \\ \hline
\multicolumn{2}{|l|}{\textbf{Total latency}} &
  \multicolumn{3}{c|}{3411 ms} &
  \multicolumn{3}{c|}{2521 ms} &
  \multicolumn{3}{c|}{1320 ms} &
  \multicolumn{3}{c|}{259 ms} \\ \hline
\end{tabular}%
}
\end{table*}


\begin{figure*}
\centering
\includegraphics[width=0.6\linewidth]{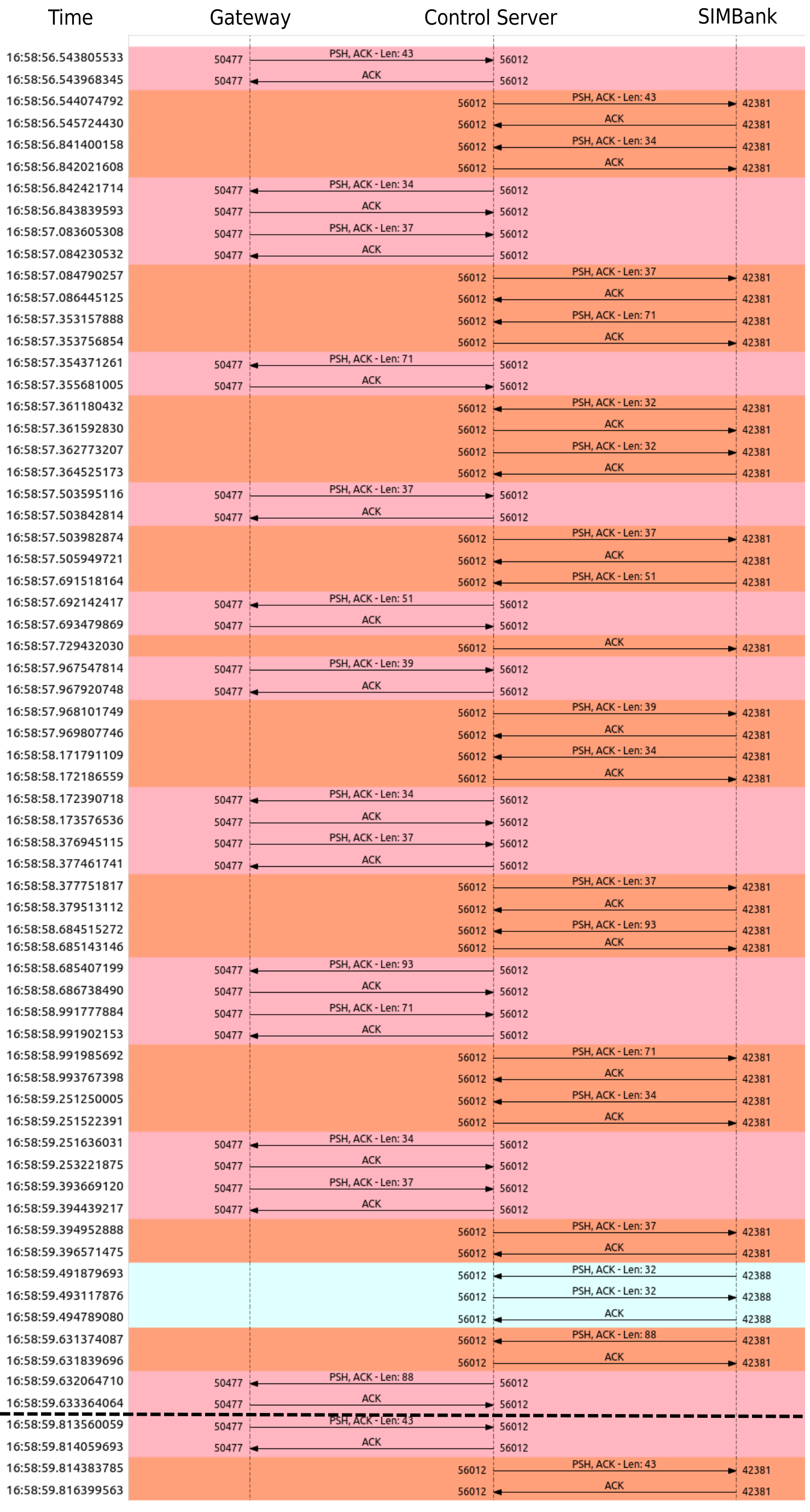}
\caption{Hybertone \simbox components TCP interactions during authentication}
\label{fig:hyb_tcp_interactions}
\end{figure*}

\begin{figure*}
\centering
\includegraphics[width=0.6\linewidth]{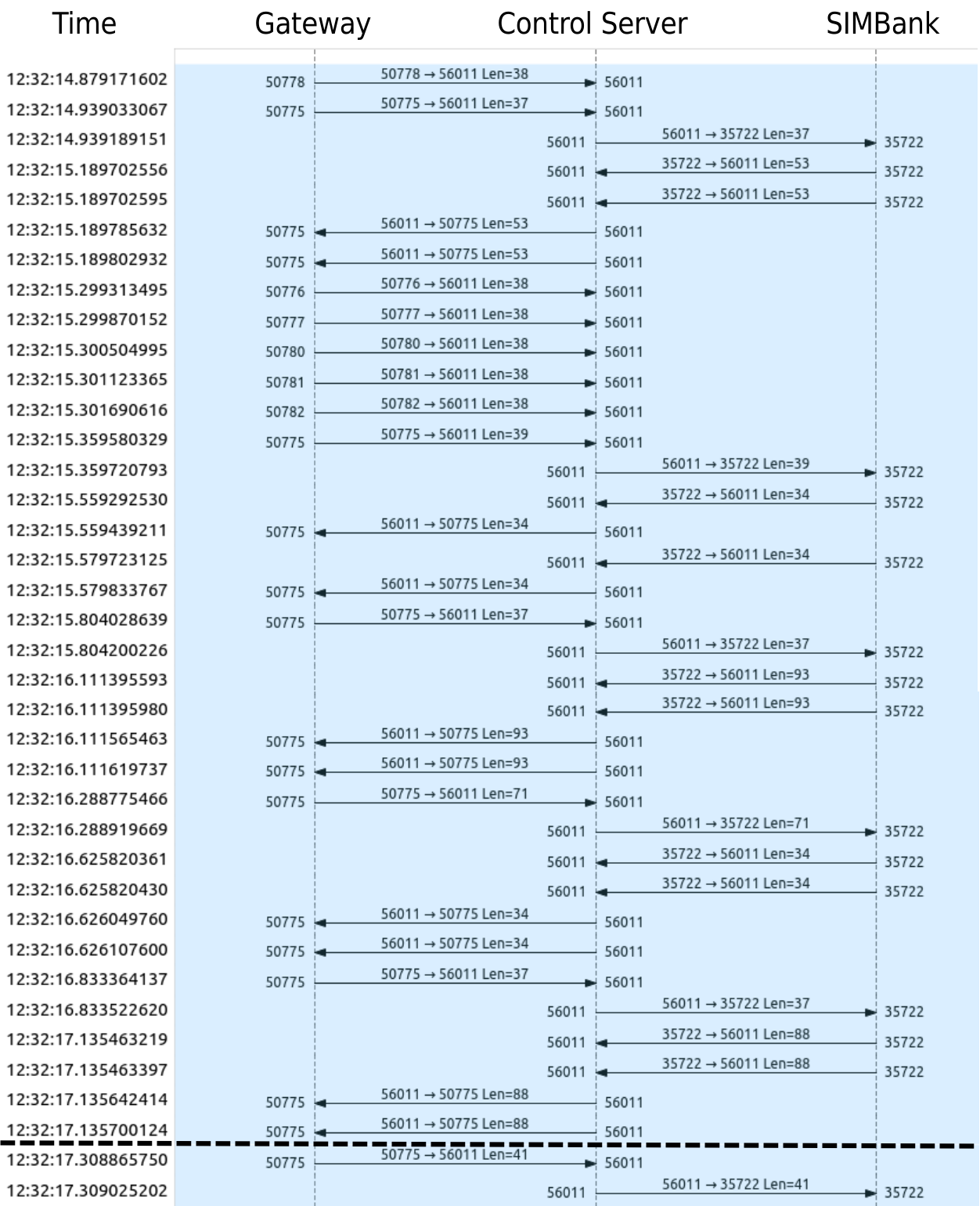}
\caption{Hybertone \simbox components UDP interactions during authentication}
\label{fig:hyb_udp_interactions}
\end{figure*}

\end{document}